\documentclass[aps,prd,reprint,groupedaddress]{revtex4-2}
\usepackage{amsmath}
\usepackage{graphicx}
\usepackage[caption=false]{subfig}

\usepackage[colorlinks=true,linkcolor=blue,citecolor=blue,urlcolor=blue]{hyperref}

\begin{document}


\title{Testing Screened Modified Gravity with Strongly Lensed Gravitational Waves}


\author{Chengsheng Mu}
\affiliation{School of Physics and Astronomy, Beijing Normal University, Beijing 100875, China}
\affiliation{Institute for Frontiers in Astronomy and Astrophysics, Beijing Normal University, Beijing 102206, China}%

\author{Shuo Cao}
\email{caoshuo@bnu.edu.cn}
\affiliation{School of Physics and Astronomy, Beijing Normal University, Beijing 100875, China}
\affiliation{Institute for Frontiers in Astronomy and Astrophysics, Beijing Normal University, Beijing 102206, China}%

\author{Shuxun Tian}
\affiliation{School of Physics and Astronomy, Beijing Normal University, Beijing 100875, China}
\affiliation{Institute for Frontiers in Astronomy and Astrophysics, Beijing Normal University, Beijing 102206, China}%

\author{Xinyue Jiang}
\affiliation{School of Physics and Astronomy, Beijing Normal University, Beijing 100875, China}
\affiliation{Institute for Frontiers in Astronomy and Astrophysics, Beijing Normal University, Beijing 102206, China}%

\author{Chenfa Zheng}
\affiliation{School of Physics and Astronomy, Beijing Normal University, Beijing 100875, China}
\affiliation{Institute for Frontiers in Astronomy and Astrophysics, Beijing Normal University, Beijing 102206, China}%

\author{Dadian Cheng}
\affiliation{School of Physics and Astronomy, Beijing Normal University, Beijing 100875, China}
\affiliation{Institute for Frontiers in Astronomy and Astrophysics, Beijing Normal University, Beijing 102206, China}%

\date{\today}

\begin{abstract}
Screening mechanisms are essential components in many modified gravity theories, which satisfy local tests of General Relativity (GR) and address cosmic acceleration on cosmological scales. The strong gravitational lensing of gravitational waves (GWs) offers a unique observational probe into cosmology and fundamental physics. In this paper, we investigate the possibility of testing screened modified gravity theories with strongly lensed gravitational waves. Specially, we develop the refined theoretical and statistical framework, in order to measure the post-Newtonian parameter $\gamma_{\text{PN}}$ in the presence of screening effects. Specially, the mass-truncated power-law and Navarro-Frenk-White (NFW) models are introduced to quantify the modified lensing potential. Our analysis also addresses the mass-sheet degeneracy (MSD) problem, by incorporating the absolute magnification and time delay measurements accessible through strongly lensed GW systems. We find that individual lensed GW system detected by next-generation GW detectors can provide stringent constraints on the PPN parameter ($\gamma_{\text{PN}}$) across different screening scales ($\Lambda$). Therefore, future measurements of strongly lensed GWs have great promise to seek departures from GR on kpc-Mpc scales, due to more precise time delay from lensed GW signals.

\end{abstract}


\maketitle

\section{Introduction}
The current cosmological paradigm is characterized by a universe undergoing accelerated expansion, a groundbreaking discovery made in 1998 through observations of distant Type Ia supernovae \cite{riessObservationalEvidenceSupernovae1998,perlmutterMeasurements42HighRedshift1999}. This unexpected phenomenon contradicted the prior assumption that gravity would universally decelerate cosmic expansion. Within the framework of General Relativity (GR), explaining this cosmic acceleration requires the existence of a repulsive energy component, now known as dark energy \cite{peeblesCosmologicalConstantDark2003}.
The cosmological constant is the simplest and most widely accepted dark energy model, initially introduced by Einstein to achieve a static universe. In the Lambda-CDM ($\Lambda$CDM) model, the cosmological constant represents a constant energy density inherent to spacetime, exerting a negative pressure that drives the accelerated expansion. While successful in fitting observational data, this model faces significant theoretical challenges, notably the ``cosmological constant problem", which concerns the vast discrepancy between its minuscule observed value and the much larger values predicted by quantum field theory \cite{carrollCosmologicalConstant2001}.

Modified gravity theories stand as a alternative explanation for the observed cosmic accelerated expansion, seeking to address this phenomenon without invoking a new dark energy component. Because tests of GR have already reached a very high precision on small scales, the viability of these theories critically hinges on the existence of screening mechanisms \cite{bakerNovelProbesProject2021}. Several modified gravity theories incorporate screening mechanisms, such as $f(R)$ gravity \cite{huModelsCosmicAcceleration2007} and scalar-tensor theories (like the Chameleon and Symmetron mechanisms) \cite{khouryChameleonCosmology2004,khouryChameleonFieldsAwaiting2004,hinterbichlerScreeningLongRangeForces2010}. In these theories, the modifications to gravity are suppressed in dense environments, typically due to environmental dependence of an additional scalar degree of freedom, which acquires a large effective mass or a weakened coupling to matter in regions of high density. Other examples include Dvali-Gabadadze-Porrati (DGP) gravity and Galileon theories, which utilize the Vainshtein mechanism to recover GR in the vicinity of massive sources through higher-derivative interactions \cite{nicolisGalileonLocalModification2009,deffayetNonperturbativeContinuityGraviton2002}. Gravitational slip refers to the inequality between the two scalar metric potentials, $\Psi$ (governing gravitational attraction) and $\Phi$ (governing spatial curvature). While GR predicts $\Psi = \Phi$ in the absence of anisotropic stress, many modified gravity theories introduce new degrees of freedom or alter gravitational interactions, leading to a non-zero slip ($\Psi \neq \Phi$) \cite{willConfrontationGeneralRelativity2014}. Detecting such a slip would thus provide compelling evidence for physics beyond standard GR and the $\Lambda$CDM model.

Gravitational lensing serves as a potent tool for probing GR and the Friedmann-Lemaître-Robertson-Walker (FLRW) at large scales \cite{2019NatSR...911608C,2019PhRvD.100b3530Q}. Numerous prior investigations have employed gravitational lensing as a means to probe GR \cite{collettPreciseExtragalacticTest2018,2020ApJ...888L..25C,2017ApJ...835...92C}. More recently, a framework for probing GR is introduced \cite{jyotiCosmicTimeSlip2019}, in the presence of screening mechanisms. Such framework incorporates a phenomenological model that posited the truncation of the gravitational slip effect within the screening scale, thereby confining its manifestation exclusively to regions beyond this scale \cite{,2022ApJ...941...16L}. However, their analysis did not fully quantify the complex degeneracies between the lens mass profile and the modified gravity parameters, potentially leading to an overestimation of the constraining power. Furthermore, a critical theoretical limitation is the behavior of the modified lensing potential near the isothermal limit. Specifically, their model exhibits a mathematical divergence (singularity) as the radial mass density slope $\gamma'$ approaches 2. Since the total mass distribution of massive elliptical galaxies—the primary targets for strong lensing—is known to be nearly isothermal ($\gamma' \approx 2$), this unphysical divergence severely restricts the applicability of their model to realistic lens systems. In this paper, we modify the previous framework \cite{jyotiCosmicTimeSlip2019} and investigate the viability of probing GR with future strongly lensed gravitational waves within a Post-Newtonian parameter framework considering screening mechanisms.

The development of gravitational wave (GW) astronomy has provided an entirely new channel to test these theories, since the direct detection of GW150914 by LIGO/Virgo \cite{collaborationGW150914FirstResults2016}. The subsequent multi-messenger observation of the binary neutron star merger GW170817 \cite{collaborationGW170817ObservationGravitational2017} placed stringent constraints on the speed of gravity \cite{collaborationGravitationalWavesGammarays2017}. Lensed gravitational waves offer exceptionally precise time delay measurements because their propagation is unimpeded by baryonic matter \cite{2021MNRAS.502L..16C,2022A&A...659L...5C}. Unlike photons, GWs are not affected by frequency-dependent delays, scattering, or absorption, providing an uncorrupted probe of the spacetime metric \cite{maggioreGravitationalWaves2008}. Furthermore, the detailed phase structure of gravitational wave waveforms allows for significantly more precise determination of arrival times compared to the often noisy electromagnetic light curves, which typically possess less precise temporal characteristics. A distinct and significant advantage of lensed gravitational waves, compared to their electromagnetic counterparts, resides in their inherent capability to ascertain the absolute magnification of each image relative to the unlensed source. This determination is facilitated by source parameters derived from gravitational wave data and precise luminosity distance measurements. Such a methodology offers a potent pathway to break the mass-sheet degeneracy (MSD) in gravitational lensing.

A critical challenge in gravitational wave astronomy is the poor sky localization of sources, which typically spans tens to hundreds of square degrees, hindering the identification of host galaxies \cite{abbottProspectsObservingLocalizing2020}. However, strong gravitational lensing offers a powerful mechanism to overcome this limitation through `lens-assisted localization' \cite{hannukselaLocalizingMergingBlack2020,zhaoLocalizationAccuracyCompact2018,nairSynergyGroundSpace2018}. If a gravitational wave event is identified as strongly lensed, its localization volume can be cross-matched with existing electromagnetic catalogs of known lenses (such as massive galaxy clusters or elliptical galaxies). Under the condition that the GW sky area is sufficiently constrained to contain only a limited number of lens candidates, this method allows for the unique identification of the host system. Consequently, the positional uncertainty is dramatically reduced from the scale of the GW error box to the sub-arcsecond precision of the optical lens galaxy. Looking ahead, next-generation ground-based detectors such as the Einstein Telescope (ET) \cite{punturoEinsteinTelescopeThirdgeneration2010,piorkowskaStrongGravitationalLensing2013} and space-based observatories like Decihertz Interferometer Gravitational Wave Observatory (DECIGO) \cite{isoyamaMultibandGravitationalWaveAstronomy2018,piorkowska-kurpasInspiralingDoubleCompact2021,kawamuraCurrentStatusSpace2020}, are poised to achieve significantly higher intrinsic localization capabilities and signal-to-noise ratios. These advancements will further refine the pre-selection of lens candidates, making the lens-assisted localization technique robust for a much larger population of events. By pinpointing the host galaxy with such high precision, we can determine the redshifts of the lens and source. This breakthrough enables the synergistic use of lensed gravitational waves and their electromagnetic signals (or properties inferred from the lens model) to jointly constrain the physical parameters of the system, providing the necessary precision to test modified gravity theories as proposed in this paper. In this paper, we address the divergence issue inherent in the power-law model by introducing a mass-truncated model, which ensures physical stability near the isothermal limit. Furthermore, we expand the theoretical scope by explicitly deriving the modified lensing potential for the Navarro-Frenk-White (NFW) density profile \cite{navarroUniversalDensityProfile1997}. Adopting a rigorous approach to error quantification, we incorporate realistic measurement uncertainties into a Bayesian inference pipeline. Finally, demonstrating the method on a representative simulated lensed gravitational wave event, we present the projected constraints on the post-Newtonian parameter $\gamma_\mathrm{PN}$ across a range of screening scales $\Lambda$.

\section{Method}
Our investigation is grounded in the theoretical framework of gravitational lensing within a perturbed cosmological spacetime. The large-scale universe is commonly described by the FLRW metric, with linear perturbations represented in the Newtonian gauge
\begin{equation}
\label{metric}
ds^2 = a^2(\tau) [-(1+2\Phi)d\tau^2 + (1-2\Psi)dx^2],
\end{equation}
where $a(\tau)$ is scale factor, $\tau$ is conformal time and $\Psi$, $\Phi$ are the conformal-Newtonian potential and longitudinal potential, respectively. These two scalar fields must be much less than one because it is the mathematical prerequisite for linearizing the Einstein field equations, ensuring that higher-order terms in the perturbative expansion can be safely neglected. This condition guarantees that the physical system under study is in the weak-gravity regime, where spacetime deviations from the smooth background are sufficiently small \cite{schneiderGravitationalLenses1992}.

Light propagation through this spacetime follows null geodesics($ds^2=0$). Solving the null geodesic equation in the \eqref{metric} reveals that the trajectory of a photon is entirely determined by the sum of two scalar fields, which defined by $\Sigma \equiv \Phi + \Psi$ \cite{schneiderGravitationalLenses1992}. In GR, these two scalar field are equal, and $\Phi$ can be solved using Poisson equation: $\nabla^2 \Phi= 4\pi G a^2\rho$. However, some modified gravity theory predicted a discrepancy between $\Phi$ and $\Psi$, known as "gravitational slip". This deviation is commonly parameterized by the Post-Newtonian parameter $\gamma_\mathrm{PN}$, defined by the relation $\Phi = \gamma_\mathrm{PN} \Psi$ \cite{willConfrontationGeneralRelativity2014}. The parameter $\gamma_\mathrm{PN}$ quantifies the degree to which space is curved by mass, differing from unity in various modified gravity theories. When screening mechanisms are considered, GR is recovered within the screening scale, and the corresponding post-Newtonian parameter $\gamma_\mathrm{PN}$ reverts to unity. One method for testing GR within the post-Newtonian framework, while incorporating screening effects, is to approximate this rapid reversion to GR within the screening radius using a step function \cite{jyotiCosmicTimeSlip2019}. 
For a spherically symmetric mass distribution $\rho(r)$, the corresponding $\Sigma(r)$ can be modeled as
\begin{equation}
    \Sigma(r)=[2+(\gamma_{\mathrm{PN}}-1)\Theta(r-\Lambda)]\Phi(r)
\end{equation}
where $\Theta$ is Heaviside step function and $\Lambda$ is screening radius.

The trajectory of null geodesics is entirely determined by the sum of two scalar field. Consequently, observables in gravitational lensing are influenced by $\Sigma(r)$. This makes gravitational lensing an appropriate tool for testing gravity. From the metric, the conformal time interval $d\tau$ for a light ray traversing a spatial path element $dl$ can be approximated as
\begin{equation}    
d\tau \approx (1 - \Phi - \Psi) dl.
\end{equation}
The total travel time for a photon from a distant source to an observer is obtained by integrating this expression along the light's path. The theoretical framework for gravitational lensing can be elegantly formulated through Fermat's Principle, which states that the actual path taken by light between two points is one for which the travel time is an extremum (i.e., $\delta \mathcal{T} = 0$, where $\mathcal{T}$ is the travel time function).
\begin{equation}
    \mathcal{T}=c^{-1}\int\left(1-{\frac{\Sigma(r)}{c^{2}}}\right)d l=c^{-1}l-c^{-3}\int \Sigma(r) d l
\end{equation}
Utilizing the thin-lens approximation, we can simplify the integral of the second term on the right-hand side of the aforementioned expression as
\begin{equation}
c^{-1} \frac{D_s D_l}{D_{ls}} (\theta-\beta)^2 - 2c^{-3}\int_{I}^{S}d r\frac{r,\Sigma(r)}{\sqrt{r^{2}-D_{l}^{2}\theta^{2}}},
\end{equation}
where $I$ denotes the image position and $S$ represents the source position. The evaluation of this integral provides the lensing potential, a fundamental quantity in gravitational lensing theory, given by
\begin{equation}
\hat{\psi}(\theta) = 2c^{-3}\int_{D_l \theta}^{S}d r\frac{r,\Sigma(r)}{\sqrt{r^{2}-D_{l}^{2}\theta^{2}}}.
\end{equation}
Given the step-like screening effect model discussed above, the integral representing the lensing potential can be split into two parts
\begin{equation}
    \hat{\psi}=\hat{\psi}_{\mathrm{GR}}+(\gamma_{\mathrm{PN}}-1)\Delta\hat{\psi}.
\end{equation}
In this expression, $\hat{\psi_{GR}}$ refers to the GR term, i.e., the lensing potential in GR, and 
\begin{equation}
\Delta\hat{\psi}(\theta)=2c^{-3}\int_{\Lambda}^{S}d r\frac{r\,\Phi(r)}{\sqrt{r^{2}-D_{l}^{2}\theta^{2}}}
\end{equation}
is the correction term due to screening effect. Directly integrating the above expression yields a very large value. However, for gravitational lensing observables such as magnification and time delay, only the difference in lensing potential at the $\theta$ positions of different images and the derivative of the lensing potential with respect to $\theta$ are relevant. Therefore, it is more common to extract the $\theta$-dependent terms from the result of the integral above, which defines the lensing potential typically in the literature.

As mentioned in the Introduction, the objective of our work is to constrain $\gamma_\mathrm{PN}$ at various values of the screening scale $\Lambda$, utilizing time delay and magnification observables from future gravitational wave events with electromagnetic counterparts. We acknowledge that the measurement of the Hubble constant is currently a subject of significant debate due to the "Hubble tension" \cite{verdeTensionsEarlyLate2019,valentinoCosmologyIntertwinedII2021}—a persistent and statistically significant discrepancy between early-universe estimates derived from the Cosmic Microwave Background \cite{collaborationPlanck2018Results2020} and late-universe measurements obtained through local distance ladders \cite{riessExpansionUniverseFaster2019}. In this paper, to maintain a consistent baseline amidst this controversy, cosmological distances are computed assuming a fiducial flat $\Lambda$CDM cosmology with a Hubble constant $H_0 = 70 \text{ km s}^{-1} \text{ Mpc}^{-1}$ and a matter density parameter $\Omega_m = 0.3$.

To accomplish the objective of constraining $\gamma_\mathrm{PN}$ for a specified value of $\Lambda$, it is important to quantify the sensitivity of gravitational lensing observables to the model parameters. In this section, we delineate the dependencies of time delay, image positions, magnification, and velocity dispersion on the model parameters, as well as the methodology for determining the posterior distribution of $\gamma_\mathrm{PN}$ via the Markov Chain Monte Carlo (MCMC) approach.
\subsection{Modified lens potential}

\subsubsection{Power-law model}
In principle, any spherically symmetric lens model can be used to calculate the modified lensing potential within the framework described above. For a power-law model, the Newtonian gravitational potential derived from the Poisson equation takes the form
$$ \Phi(r)={\frac{4\pi G\rho_{0}r_{0}^{\gamma}}{(\gamma^{\prime}-3)(\gamma^{\prime}-2)}}\,r^{2-\gamma^{\prime}}, $$
where $1.5 < \gamma' < 2.5$.

This form implicitly implies that for $\gamma' < 2$, the gravitational potential diverges at infinity, consequently leading to the scalar field $\Phi$ failing to satisfy the condition of being much smaller than $1$ during the integral calculation of the lensing potential. This problem becomes particularly pronounced when considering screening mechanisms; it would lead to the correction potential diverging as $\gamma' \to 2$ if calculated directly following the aforementioned framework. This is absurd. 

For power-law models, the core solution to this divergence issue is to introduce a cutoff radius, postulating that the dark matter halo density rapidly decreases to zero or near zero beyond a certain finite radius. This is physically reasonable, as dark matter halos are subject to tidal forces. In this paper, we consider a sharp cutoff model, where the mass distribution of a spherically symmetric lens galaxy can be written as
\begin{equation}
    \rho_m(r) = \begin{cases}
    \rho(r) &  r < r_m \\
    0 &  r >r_m .
\end{cases}
\end{equation}
This step-like model ensures a finite total mass for the lens galaxy by truncating the mass distribution at $r=r_m$, thereby resolving the aforementioned divergence at infinity. In the context of our study, the screening scale is assumed to be smaller than the cutoff radius of the galaxy's mass distribution. Consequently, we only consider the scenario where $r_m >\Lambda$ in this work.

Upon considering a truncated mass distribution model, the Newtonian potential of the lens galaxy can be derived by solving the Poisson equation. For $r \ge r_m$, the potential is given by
$$ \Phi(r) = \frac{4\pi G\rho_0r_0}{3-\gamma'}r_m^{3-\gamma'}\frac{1}{r}. $$
For $r < r_m$, the potential takes the form
$$ \Phi(r) = \frac{4\pi G\rho_0r_0}{3-\gamma'}r_m^{2-\gamma'} + \frac{4\pi G\rho_0r_0}{(3-\gamma')(2-\gamma')}(r^{2-\gamma'} - r_m^{2-\gamma'}).$$
Though the expressions above encounter a denominator-zero problem when $\gamma' = 2$ (corresponding to the Singular Isothermal Sphere (SIS) model), it can be proven that the Newtonian potential as $\gamma'$ approaches $2$ is fully consistent with the result obtained directly from the SIS model.

The corrected lens potential can be calculated based on this Newtonian potential using the aforementioned most direct method.
The integral for $\Delta\psi(\theta)$ is expressed as
$$ \Delta\hat{\psi}(\theta)=\frac{8\pi G\rho_{0}\,r_{0}^{\gamma'}}{(\gamma^{\prime}-3)(\gamma^{\prime}-2)}\int_{\Lambda}^{r_{m}}\frac{r^{3-\gamma^{\prime}}}{\sqrt{r^{2}-D_{L}^{2}\theta^{2}}}\,d r $$
This integral can be analytically evaluated by a methodical decomposition into three distinct parts. Ultimately, the result of this integration can be expressed as a sum of three terms
$$ \Delta \hat{\psi} = \Delta \hat{\psi_1} + \Delta \hat{\psi_2} +\Delta \hat{\psi_3} $$
After discarding the terms in the integral result that are independent of $\theta$, the result for each component is as follows
\begin{widetext}

\begin{equation}
 \Delta\hat{\psi_1}(\theta)=\frac{\theta^{3-\gamma'}\theta_\mathrm{E,GR}^{\gamma'-1}}{2-\gamma'}\frac{D_LD_S}{D_{LS}\sqrt{\pi}}\frac{\Gamma(\frac{\gamma'}{2})}{\Gamma(\frac{\gamma '-1}{2})} \left[(1-\frac{D_L^2\theta^2}{r^2})^{\frac{1}{2}}\,_2F_1\left(\frac{1}{2},\frac{5-\gamma'}{2};\frac{3}{2},1-\frac{D^2_L\theta^2}{r^2}\right)\right]_{\Lambda}^{r_m}
\end{equation}
\begin{equation}
 \Delta\hat{\psi_2}(\theta) = \left({\frac{r_{m}}{D_{L}}}\right)^{3-\gamma^{\prime}}\theta_{E, \mathrm{GR}}^{\gamma'-1}{\frac{D_{S}D_{L}}{D_{L S}}}\ln\left({r_m +\sqrt{r_m^2 - D_L^2  \theta^2}}\right)\frac{\Gamma(\frac{\gamma'}{2})}{2\sqrt{\pi}\Gamma(\frac{\gamma '-1}{2})}
\end{equation}
\begin{equation}
 \Delta\hat{\psi_3}(\theta) = \frac{3-\gamma'}{\gamma'-2} \left(\frac{r_m}{D_L}\right)^{2-\gamma'} \frac{D_S}{D_{LS}} \theta_{\mathrm{E,GR}}^{\gamma'-1} \frac{\Gamma(\frac{\gamma'}{2})}{\sqrt{\pi}\Gamma(\frac{\gamma '-1}{2})}\left(\sqrt{r_m^2 - D_L^2 \theta^2} - \sqrt{\Lambda^2 - D_L^2 \theta^2}\right)
\end{equation}
\end{widetext}
Here, $\theta_\mathrm{E,GR}$ represents the Einstein angle for a spherically symmetric lens in GR, which for a power-law model, takes the form
$$\theta_\mathrm{E,GR}^{\gamma'-1} = \frac{2\rho_0 r_0^{\gamma'}\sqrt{\pi}}{3-\gamma'}D_l^{1-\gamma'}\frac{\Gamma(\frac{\gamma'-1}{2})}{\Gamma(\frac{\gamma'}{2})}\frac{4\pi GD_lD_{ls}}{c^2D_s}.$$

The Einstein radius $\theta_{E,\mathrm{GR}}$ serves as a primary indicator of the lens mass distribution, defining the characteristic angular scale where the average surface mass density equals the critical surface density for lensing. Within the framework of the power-law model, $\theta_{E,\mathrm{GR}}$ acts as a fundamental normalization parameter that directly links the observed image configurations to the total projected mass enclosed within the Einstein ring. To characterize the physical extent of the lens halo, we identify the truncation radius $r_m$ with the virial radius $R_{\text{vir}}$ of the lens galaxy. The virial radius is defined such that the mean density within this sphere equals 200 times the critical density of the universe, $\rho_{\text{crit}}(z_l)$, at the lens redshift \cite{gunnInfallMatterClusters1972,navarroUniversalDensityProfile1997}. By relating the density normalization to the observable $\theta_{E,\mathrm{GR}}$, we derive the analytical expression for the virial radius as
\begin{equation}
R_{\text{vir}} = \left[ \frac{c^2 D_s D_l^{\gamma'-2} \theta_{E,\mathrm{GR}}^{\gamma'-1}}{200 H(z_l)^2 D_{ls} \sqrt{\pi}} \frac{\Gamma(\gamma'/2)}{\Gamma((\gamma'-1)/2)} \right]^{1/\gamma'}
\end{equation}
where $D_l$, $D_s$, and $D_{ls}$ denote the angular diameter distances to the lens, to the source, and between the lens and source, respectively, and $H(z_l)$ represents the Hubble parameter at the lens redshift.

\subsubsection{NFW model}
The Navarro-Frenk-White profile is a widely adopted analytical model for the density distribution of dark matter halos, derived from extensive N-body simulations of hierarchical structure formation in a $\Lambda$CDM universe \cite{navarroUniversalDensityProfile1997}. It describes the mass density $\rho(r)$ as a function of radius $r$ from the halo center, given by
$$\rho(r) = \frac{\rho_0}{\left(r/r_s\right)(1 + r/r_s)^2}$$
where $\rho_0$ is a characteristic density, and $r_s$ is a scale radius. This profile is characterized by a "cuspy" inner region ($\rho(r) \propto r^{-1}$ as $r \to 0$) and a steeper decline in the outer regions ($\rho(r) \propto r^{-3}$ as $r \to \infty$). The NFW model has been highly successful in reproducing the averaged density profiles of simulated dark matter halos across a wide range of masses and redshifts, making it an indispensable tool for interpreting astrophysical observations, including gravitational lensing.

For the NFW model, the correction term to the lensing potential, $\Delta\psi$, can be directly obtained from the following integral expression
$$ \Delta \hat{\psi} = -4\pi G \rho_0 r_s^3 \int^S_\Lambda \frac{\ln(1+r/r_s)}{\sqrt{r^2-D_l^2\theta^2}} dr $$
Considering the limit as $S \to \infty$ and discarding terms independent of $\theta$, the form of the lensing potential correction term is found to be
\begin{equation}
    \Delta \hat{\psi} = -4\pi G \rho_0 r_s^3\left[P - F\left(u_\Lambda, k, \gamma_k\right)\right]
\end{equation} 
In this expression, $P$ and $F$ are defined as follows, with
$$ P = -\frac{1}{2}(\log a)^{2}+(\log2+\log r_{s})\log a-\gamma_{k}^{2} $$
$$ F(u, k, \gamma_k) = u \log\left(\frac{k}{2}\right) - \frac{u^2}{2} - 2 \text{Re}\left[\text{Li}_2\left(-e^{u + i\gamma_k}\right)\right] $$
where $\text{Li}_2$ is the dilogarithm function, and the other variables are defined as
$a = D_l\theta$,
$k = a/r_s$,
$\gamma_k = arccos\left(1/k\right)$,
$u_\Lambda = acosh\left(\Lambda/a\right)$.
\subsection{Time delay}
Given the lensing potential, the Fermat potential in gravitational lensing is defined by the following expression
\begin{equation}
     \hat{\phi}(\theta,\beta)=\frac{{D_{l}}D_{s}}{2D_{ls}}\left(\theta-\beta\right)^{2}-\hat{\psi}(\xi)
\end{equation}
In this formulation, $\theta$ refers to the image position and $\beta$ to the source position. It is important to note that for a given lensing event, the source position is an unobservable, fixed quantity. The extrema of the Fermat potential correspond to the image positions. The time delay between two images is precisely the difference in their respective Fermat potentials. Therefore, the expression for the time delay between image $i$ and image $j$ is given by
\begin{equation}
    \Delta t_{ij} = \frac{{D_{l}}D_{s}}{2D_{ls}}\left[ (\theta_i-\beta)^2-(\theta_j-\beta)^2\right]+\hat{\psi_i}-\hat{\psi_j}.
\end{equation}

\subsection{Magnification}
Magnification is another crucial observable in gravitational lensing. Given the explicit expression for the lensing potential, obtaining the magnification formula is, in principle, straightforward \cite{schneiderGravitationalLenses1992,schneiderGravitationalLensingStrong2006,narayanLecturesGravitationalLensing1997}.
The lens equation can be viewed as a transformation from the source plane to the image plane, with magnification equivalent to the Jacobian determinant of this transformation \cite{schneiderGravitationalLenses1992,schneiderGravitationalLensingStrong2006}.

For an axisymmetric lens, the (reduced) deflection angle is given by the angular gradient of the lensing potential. With our definition of $\hat{\psi}(\theta)$, one has
\begin{equation}
\alpha(\theta)= \frac{D_{ls}}{D_l D_s} \nabla_{\theta} \hat{\psi}({\theta}).
\end{equation}
The magnification rate $\mu$ of a spherically symmetric lens can be expressed by
\begin{equation}
\label{Mag}
    \mu(\theta) = \frac{1}{\left(1 - \frac{\alpha(\theta)}{\theta}\right) \left(1 - \frac{d\alpha(\theta)}{d\theta}\right)}.
\end{equation}
The observational uncertainty associated with the lensed gravitational wave magnification can be estimated through the application of the error propagation formula to the luminosity distance uncertainty.
In general, the luminosity distance error budget contains an instrumental component and an (astrophysical) weak-lensing component,
\begin{equation}
     \sigma_{d_{L}}=\sqrt{(\sigma_{d_{L}}^{\mathrm{instr}})^{2}\,+\,(\sigma_{d_{L}}^{\mathrm{lens}})^{2}.}
\end{equation}
In this work, we focus on the regime most relevant for third-generation detectors and confidently identified lens systems. Here $\sigma_{d_{L}}^{\mathrm{instr}}$ denotes the uncertainty in the luminosity-distance measurement arising from the detector noise and the statistical limitations in determining the gravitational-wave source parameters (in particular the binary inclination angle), whereas $\sigma_{d_{L}}^{\mathrm{lens}}$ denotes the uncertainty induced by weak gravitational lensing from large-scale structures along the propagation path.
For the latter, we adopt the commonly used approximation \cite{holzUsingGravitationalwaveStandard2005,zhaoDeterminationDarkEnergy2011}
\begin{equation}
\sigma_{d_L}^{\mathrm{lens}} \simeq 0.05\, z\, d_L, \qquad (z\lesssim 1),
\end{equation}
and in the remainder of this paper we will not further discuss the detailed modeling of $\sigma_{d_L}$ beyond this prescription.

The luminosity distance inferred from a lensed gravitational-wave observation under the assumption of no lensing, $d_L^{\mathrm{obs}}$, relates to the true (unlensed) luminosity distance $d_L^{\mathrm{true}}$ and the gravitational magnification $\mu$ by
\begin{equation}
    d_{L}^{\mathrm{obs}}=\frac{d_{L}^{\mathrm{true}}}{\sqrt{\mu}}.
\end{equation}
This implies that $\mu = (d_L^{\mathrm{true}} / d_L^{\mathrm{obs}})^2$. Using standard error propagation, the relative uncertainty on magnification can be written as
\begin{equation}
\left(\frac{\Delta \mu}{\mu}\right)^2 = 4\left[\left(\frac{\Delta d_L^{\mathrm{obs}}}{d_L^{\mathrm{obs}}}\right)^2+\left(\frac{\Delta d_L^{\mathrm{true}}}{d_L^{\mathrm{true}}}\right)^2\right].
\end{equation}
If $d_L^{\mathrm{true}}$ is treated as fixed (or its uncertainty is negligible compared to $\Delta d_L^{\mathrm{obs}}$), this reduces to $\Delta \mu/\mu \simeq 2\,\Delta d_L^{\mathrm{obs}}/d_L^{\mathrm{obs}}$. In this work, we neglect the uncertainty of $d_L^{\mathrm{true}}$, which is primarily induced by cosmological parameter uncertainties (e.g., $H_0$).
Furthermore, lensed gravitational waves with electromagnetic counterparts offer a unique advantage by allowing the measurement of absolute magnification. This capability is crucial for breaking the MSD.
\subsection{Velocity dispersion}
In this section, we follow the Ref. \cite{koopmansGravitationalLensingStellar2006} to compute the velocity dispersion. Velocity dispersion depends only on the 00 component of the metric $g_{\mu\nu}$; in the weak-field, non-relativistic limit it does not directly depend on the Post-Newtonian parameter that controls $g_{ij}$. To incorporate velocity dispersion observations into parameter estimation, the relationship between velocity dispersion and mass distribution must be calculated.

For a spherically symmetric lens galaxy, the line-of-sight (LOS) velocity dispersion is derived from its total enclosed mass profile $M(r)$ (including both baryons and dark matter) via the spherical Jeans equation \cite{2015ApJ...806..185C}. We assume isotropic stellar orbits, i.e., a velocity anisotropy parameter $\beta_{\mathrm{ani}}=0$, and describe the tracer (stellar) distribution with $\rho_*(r)$ \cite{2016MNRAS.461.2192C}. In general, the tracer distribution $\rho_*(r)$ does not necessarily follow the galaxy's total mass distribution. Therefore, $\rho_*(r)$ should be modeled independently and constrained with observational data (e.g., surface brightness profiles from imaging). For simplicity, we adopt a simplified assumption that the tracer distribution follows the mass distribution in this work.

The 3D radial velocity dispersion $\sigma_r^2(r)$ then satisfies
\begin{equation} 
\rho_*(r)\sigma_r^2(r) = \int_r^\infty \rho_*(r') \frac{GM(r')}{r'^2} dr'.
\end{equation}
This is then projected onto the line of sight to yield $\sigma_{\mathrm{LOS}}^2(R)$ at projected radius $R$.
$$\Sigma_*(R)\sigma_{\mathrm{LOS}}^2(R) = 2 \int_R^\infty \rho_*(r)\sigma_r^2(r) \frac{r}{\sqrt{r^2-R^2}} dr,$$
where $\Sigma_*(R)$ is the projected stellar surface density.

Finally, aperture correction for an observed aperture radius $R_{ap}$ is calculated by averaging 
\begin{equation}
\langle \sigma_{\mathrm{LOS}}^2 \rangle_{ap} = \frac{\int_0^{R_{ap}} \Sigma_*(R)\sigma_{\mathrm{LOS}}^2(R) 2\pi R dR}{\int_0^{R_{ap}} \Sigma_*(R) 2\pi R dR}.
\end{equation}
Different mass distribution models correspond to different velocity dispersions. The observed velocity dispersion provides dynamical information that helps constrain lens mass-related parameters (e.g., $\theta_{E,\mathrm{GR}}$).
\subsection{Mass-sheet Degeneracy}
Constrain $\gamma_\mathrm{PN}$ using gravitation
The mass-sheet degeneracy is a fundamental ambiguity inherent in gravitational lensing observations. It states that for any given lens mass distribution that produces an observed set of lensed images, one can add a uniform sheet of mass to the lens plane and simultaneously rescale the lens mass and source positions such that the observed image positions and shapes remain unchanged \cite{schneiderSourcepositionTransformationApproximate2014, schneiderMasssheetDegeneracyPowerlaw2013}.

The total lensing potential is a simple linear superposition of the potentials contributed by these two distinct physical components
\begin{equation}
    \psi_{\text{total,GR}}(\boldsymbol{\theta}) = \psi_\mathrm{GR}(\boldsymbol{\theta} ) + \psi_{\text{sheet}}(\boldsymbol{\theta}),
\end{equation}
where, $\psi_{\text{sheet}}(\boldsymbol{\theta})$ denotes the potential produced by a physical mass sheet. For a mass sheet characterized by constant convergence, its potential is expressed as
\begin{equation}
\psi_{\text{sheet}}(\boldsymbol{\theta}) = \frac{1}{2}\kappa_{\text{sheet}} |\boldsymbol{\theta}|^2
\end{equation}
The MSD fundamentally limits gravitational lensing analyses by allowing for an arbitrary scaling of the lens mass distribution while preserving relative image positions and magnifications. This ambiguity can be broken (or significantly mitigated) by incorporating information sensitive to the absolute mass scale. In particular, the H0LiCOW time-delay cosmography program demonstrates that combining high-resolution lensing imaging with stellar kinematics (velocity dispersion) provides a dynamical constraint on the lens mass normalization and profile, thereby helping to control the MSD and related degeneracies in practical lens modeling \cite{suyuH0LiCOW$H_0$Lenses2017,rusuH0LiCOWXIILens2020,wongH0LiCOWIVLens2017}.

In our framework, we similarly incorporate velocity dispersion measurements to provide an independent dynamical anchor on the total enclosed mass of the lens galaxy. Furthermore, lensed gravitational waves with electromagnetic counterparts offer an additional probe that is typically unavailable in electromagnetic-only lensing: the absolute magnification of each image (e.g., via $D_L^{\mathrm{obs}} = D_L^{\mathrm{true}} / \sqrt{\mu}$). This observable directly constrains the external convergence (mass-sheet parameter) $\kappa_{\text{sheet}}$ through the mass-sheet potential $\psi_{\text{sheet}}(\boldsymbol{\theta}) = \frac{1}{2}\kappa_{\text{sheet}} |\boldsymbol{\theta}|^2$, because the absolute magnification is scaled by a factor of $(1-\kappa_{\text{sheet}})^{-2}$. Consequently, introducing absolute magnification as an extra probe tightens the posterior on $\kappa_{\text{sheet}}$ and improves the overall robustness of the inference against the MSD \cite{suyuH0LiCOW$H_0$Lenses2017,schneiderMasssheetDegeneracyPowerlaw2013}. The physical origin of the MSD is the external convergence from the large-scale environment of the lens, which acts as a near-uniform sheet of mass across the strong-lensing region. This environment includes the host galaxy group or cluster, neighboring line-of-sight galaxies, and the cosmic web filaments and voids. In this work, we consider that these effects occur outside the screening scale of modified gravity, and are thus unaffected by the screening mechanism.

To incorporate this effect and evaluate our method's constraining power on model parameters, we adopt the following lens potential as
\begin{equation}
    \hat{\psi}_\mathrm{total}(\boldsymbol{\theta}) = \hat{\psi}_\mathrm{GR}(\boldsymbol{\theta})+\Delta\hat{\psi}(\boldsymbol{\theta})+ \frac{1+\gamma_\mathrm{PN}}{2}\hat{\psi}_\mathrm{sheet}(\boldsymbol{\theta}).
\end{equation}

\section{Results and discussion}
For the problem at hand, Bayes' theorem stipulates that the posterior probability density function (PDF) is expressed as
\begin{equation}
    \mathcal{P}(\boldsymbol{\theta}|\boldsymbol{D})\propto\mathcal{L}( \boldsymbol{D}, \boldsymbol{\sigma}|\boldsymbol{\theta})\mathcal{\pi}(\boldsymbol{\theta}) 
\end{equation}
Our analysis employs a Bayesian inference framework to constrain $\gamma_\mathrm{PN}$ and lens parameters. Specifically, for a power-law model, the pertinent lensing parameters encompass the source position $\beta$, the Einstein angle in GR $\theta_\mathrm{E,GR}$, and the power-law index $\gamma'$ of the galaxy.

We employ a multiplicative Gaussian likelihood approach to evaluate the model parameters. The likelihood function, $\mathcal{L}(\boldsymbol{\theta} | \boldsymbol{D}, \boldsymbol{\sigma}) $, for a set of observations $\boldsymbol{D} = \{D_1, D_2, \dots, D_N\}$ given a model $\mathbf{M}(\boldsymbol{\theta}) = \{M_1(\boldsymbol{\theta}), M_2(\boldsymbol{\theta}), \dots, M_N(\boldsymbol{\theta})\}$ and corresponding uncertainties $\boldsymbol{\sigma} = \{\sigma_1, \sigma_2, \dots, \sigma_N\}$, is expressed as
\begin{equation}
\mathcal{L}(\boldsymbol{\theta} | \boldsymbol{D}, \boldsymbol{\sigma}) = \prod_{i=1}^{N} \frac{1}{\sqrt{2\pi\sigma_i^2}} \exp\left(-\frac{(D_i - M_i(\boldsymbol{\theta}))^2}{2\sigma_i^2}\right).
\end{equation}
Following this, we perform sampling of the posterior probability distribution function (PDF) for the nonlinear lens parameters using the MCMC method. The `emcee' Python package is utilized to achieve efficient sampling of this posterior distribution \cite{foreman-mackeyEmceeMCMCHammer2013}.

The exceedingly small measurement uncertainties associated with time delay measurements present significant challenges for MCMC sampling. A proper strategy to solve this difficulty involves treating the time delay as a stringent constraint, thereby enabling the reduction of the parameter space dimensionality prior to sampling. During the sampling process, this dimensionality reduction is executed by a numerical solver. To operationalize this approach, a suitable parameter must be selected for numerical evaluation. For the time-delay expression in this study, $\gamma_\mathrm{PN}$ is chosen owing to its straightforward dependence, which makes it the preferred choice to circumvent numerical ambiguities arising from multiple solutions. 

\begin{figure*}
    \centering
    \includegraphics[width=0.7\linewidth]{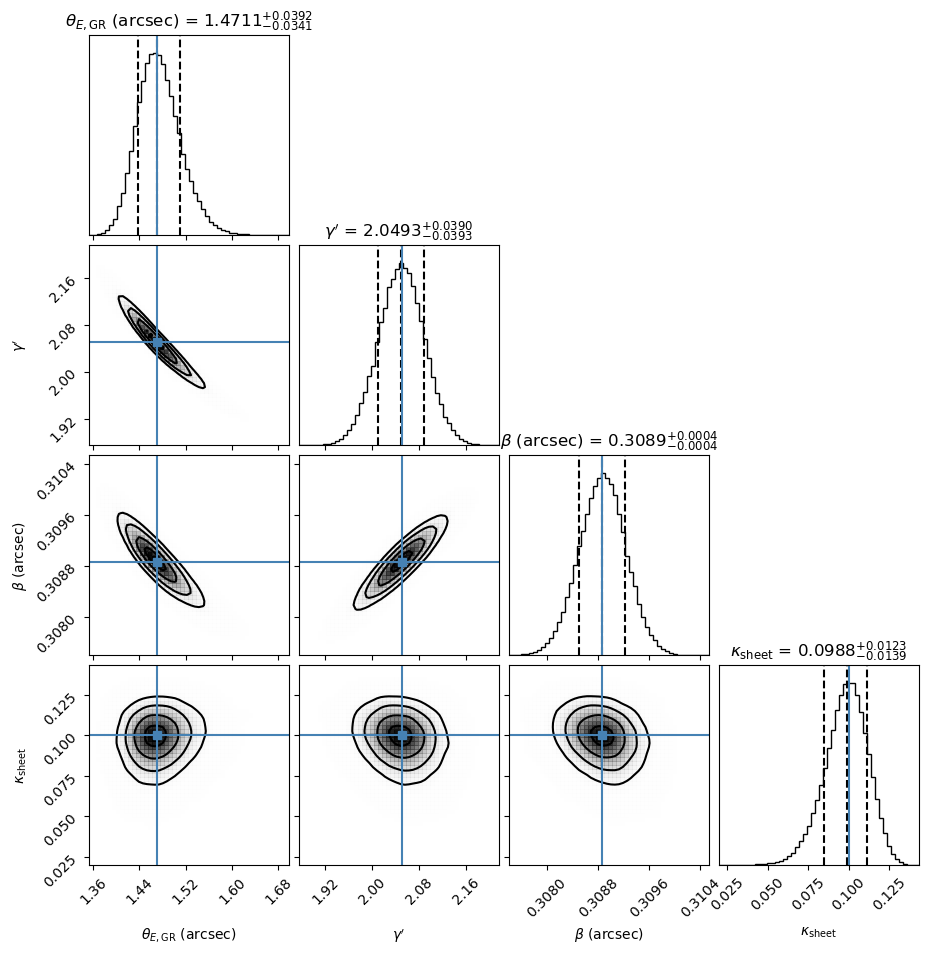}
    \caption{Posterior probability distributions of the model parameters for Event I. The diagonal panels show the 1-D marginalized distributions with dashed lines indicating the median and the $1\sigma$ confidence intervals (16\% and 84\%). The off-diagonal panels display the 2-D joint confidence regions.}
    \label{fig:enter-label}
\end{figure*}

\begin{figure*}[htbp]
    \centering
    \begin{tabular}{@{}cc@{}}  
        \subfloat[System I, $\gamma_\mathrm{PN} = 1.0$\label{fig:sub1}]{
            \includegraphics[width=0.45\textwidth]{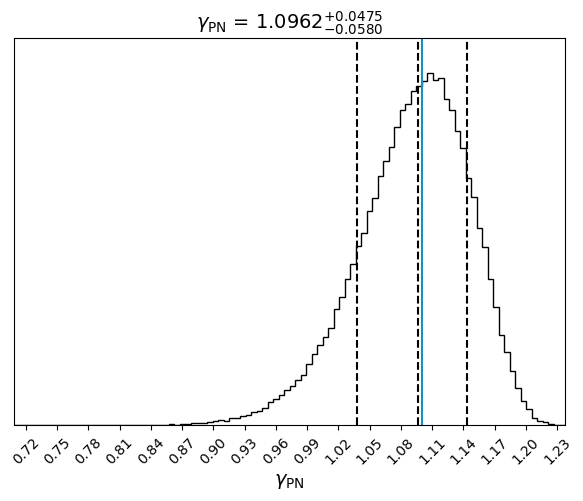}
        } &
        \subfloat[System II, $\gamma_\mathrm{PN} = 1.0$\label{fig:sub2}]{
            \includegraphics[width=0.45\textwidth]{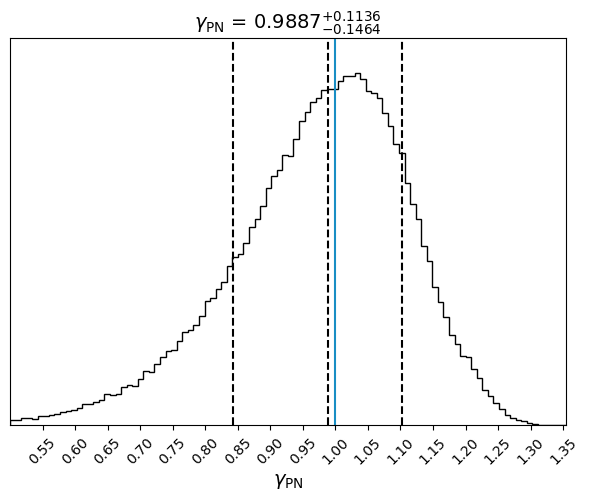}
        } \\

        \\[-10pt]

        \subfloat[System I, $\gamma_\mathrm{PN} = 1.1$\label{fig:sub3}]{
            \includegraphics[width=0.45\textwidth]{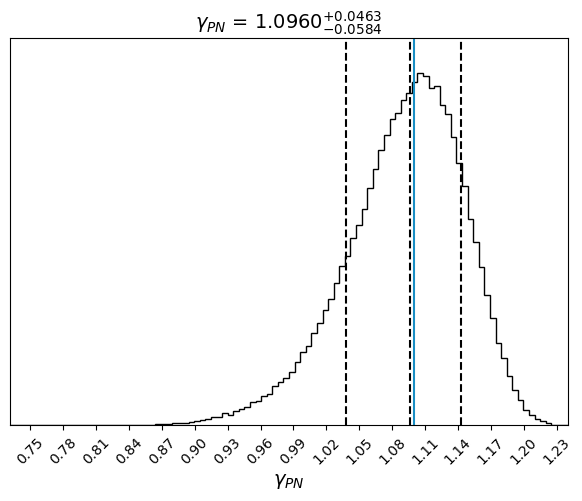}
        } &
        \subfloat[System II, $\gamma_\mathrm{PN} = 1.1$\label{fig:sub4}]{
            \includegraphics[width=0.45\textwidth]{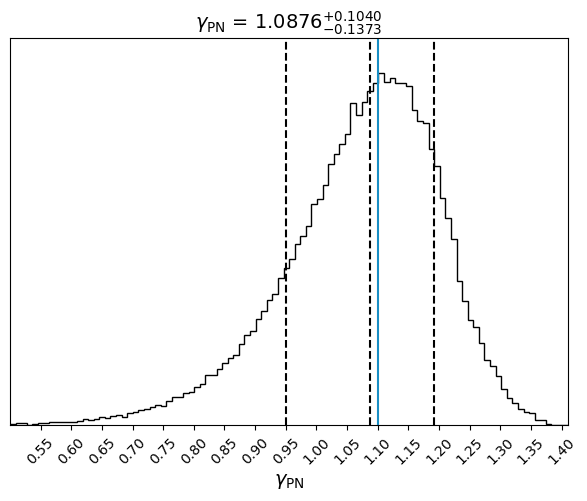}
        }
    \end{tabular}

    \caption{One-dimensional posterior probability distributions of the Post-Newtonian parameter $\gamma_\mathrm{PN}$. The left columns (a), (c) and right columns (b), (d) correspond to System I and System II, respectively. The top row shows the recovered results for an injected signal consistent with GR ($\gamma_\mathrm{PN}=1$), while the bottom row displays the results for a simulated modified gravity signal ($\gamma_\mathrm{PN}=1.1$). The solid blue lines indicate the fiducial values, and the vertical dashed lines represent the 68\% ($1\sigma$) confidence intervals.}
    \label{fig:main}
\end{figure*}

To robustly demonstrate the potential of lensed GWs as a future probe for GR, we simulate a series of gravitational-wave events lensed by a mass-truncated power-law model \cite{piorkowskaStrongGravitationalLensing2013,biesiadaStrongGravitationalLensing2014,piorkowska-kurpasInspiralingDoubleCompact2021,2019NatSR...911608C}.
In the generated suite of lensed gravitational-wave data, we arbitrarily select two cases for illustration. The parameters for Lens System I are source redshift, $z_s = 1.038$,  lens redshift, $z_l = 0.522$, source position $\beta = 0.309''$, Einstein radius in GR $\theta_{E,\mathrm{GR}} = 1.47''$.  The parameters for Lens System II are source redshift, $z_s = 0.615$,  lens redshift, $z_l = 0.346$, source position $\beta = 0.389''$, Einstein radius in GR $\theta_{E,\mathrm{GR}} = 1.09''$. The power indices for both lens events are 2.05.

The measurement uncertainty for image positions is adopted as 0.001 arcsec. Furthermore, uncertainties of approximately $5\%$ are considered for the velocity dispersion. Regarding the absolute magnification in gravitational lensing, to simplify the discussion we assume that the observed events have sufficiently high signal-to-noise ratios such that the luminosity-distance uncertainty is dominated by the weak-lensing term $\sigma_{d_L}^{\mathrm{lens}}$. Therefore, after calculation, the relative errors of the magnification for these two events were taken as 10\% and 6\% respectively. To constrain the radial mass density profile of the lens galaxy, we refer to the robust methodology framework established by the H0LiCOW collaboration \cite{suyuH0LiCOW$H_0$Lenses2017,rusuH0LiCOWXIILens2020}. Their approach precisely determines the total mass density slope, $\gamma'$, by breaking the inherent degeneracy between the mass profile shape and orbital anisotropy. This is achieved through a joint likelihood analysis that combines high-resolution lensing imaging -- which reconstructs the pixel-level surface brightness of Einstein rings and lensed arcs -- with stellar kinematic measurements (velocity dispersion). Informed by their high-precision constraints, we assign a Gaussian prior to the power-law slope $\gamma'$ in our Bayesian MCMC inference. We set the standard deviation of this prior to $\sigma_{\gamma'} = 0.04$. This value is chosen as a realistic benchmark for high-quality strong lens systems, as it aligns empirically with the absolute precision achieved for several systems within the H0LiCOW sample. Notably, analyses of lenses such as WFI 2033-4723 \cite{rusuH0LiCOWXIILens2020}, RXJ 1131-1231 \cite{suyuTwoAccurateTimedelay2013}, and HE 0435-1223 \cite{wongH0LiCOWIVLens2017} have demonstrated that measurement uncertainties on the order of $\sigma_{\gamma'} \approx 0.04$ to $0.05$ are attainable with current state-of-the-art facilities.

\begin{figure*}
    \centering
    \includegraphics[width=0.7\linewidth]{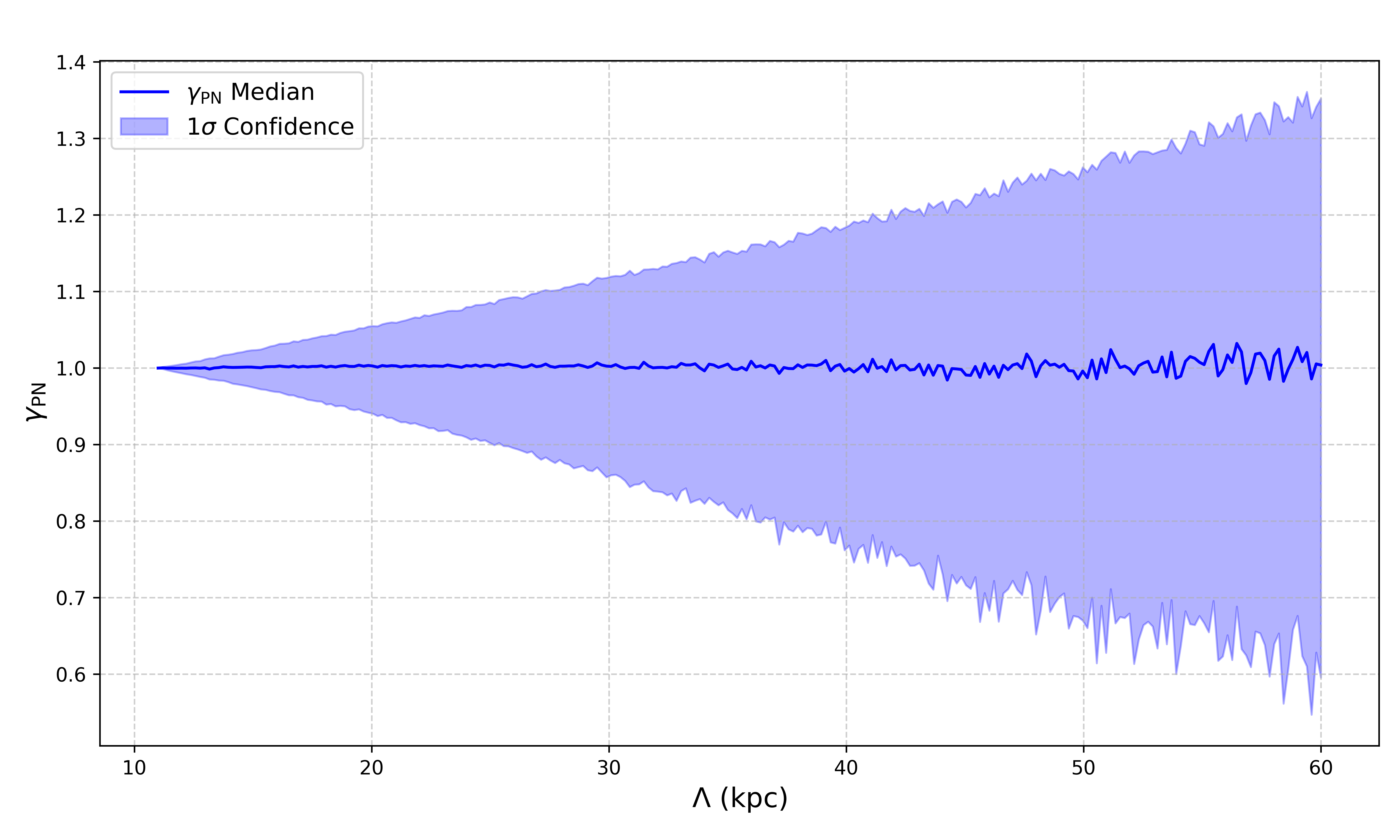}
    \caption{Constraints on the post-Newtonian parameter $\gamma_\mathrm{PN}$ as a function of the screening scale $\Lambda$ (in kpc). The results are derived from an ensemble of 250 independent MCMC simulations. The solid blue line indicates the median of the posterior distributions, while the light blue shaded region represents the $1\sigma$ (68\%) confidence interval. The widening of the confidence band indicates that the constraints on $\gamma_\mathrm{PN}$ degrade as the screening scale $\Lambda$ increases.
}
    \label{fig:enter-label2}
\end{figure*}

Based on the aforementioned framework, we calculate the observables for Lens System I and Lens System II under two distinct parameter configurations. In the first scenario, we assign $\gamma_\mathrm{PN} = 1 $ and $\Lambda = 20$ kpc to generate the observed data for Systems I and II, respectively. In the second scenario, a shifted value of $\gamma_\mathrm{PN} = 1.1$(with $\Lambda = 20$ kpc) is adopted to obtain the observables for Systems I and II, respectively. Subsequently, MCMC analysis is performed on these four datasets to reconstruct the respective posterior distributions of $\gamma_\mathrm{PN}$.  Figure~\ref{fig:enter-label} illustrates the joint posterior distributions of the lens parameters derived from the MCMC sampling for System I. Figure~\ref{fig:main} presents a comparison of the reconstructed $\gamma_\mathrm{PN}$ distributions for the two representative systems under different injected values. To further explore how the screening scale affects the constraining power on $\gamma_{\mathrm{PN}}$, we conduct 250 independent MCMC analyses by uniformly sampling $\Lambda$ within the interval $[11, 60]\,\mathrm{kpc}$. The resulting $1\text{-}\sigma$ uncertainties of $\gamma_{\mathrm{PN}}$ for System I are plotted against different $\Lambda$ values in Figure~\ref{fig:enter-label2}.

It is found that constraints on $\gamma_\mathrm{PN}$ are maximized by systems combining a large Einstein radius ($R_E$) with a low source redshift ($z_s$). Physically, low $z_s$ ensures high SNR for gravitational waves, yielding precise absolute magnification measurements essential for breaking the MSD. Simultaneously, massive lenses with large physical $R_E$ increase the probability that images form in the unscreened regime ($r \gtrsim \Lambda$), thereby maximizing the exposure to modified gravity effects.  This paper focuses on screening scales within the 10–100 kpc range, a regime where the fifth force within the galaxy is partially screened. This approach provides a viable framework for investigating modified gravity on larger scales while ensuring that the model remains consistent with stringent Solar System constraints.

An inherent degeneracy is observed between the screening scale $\Lambda$ and the interaction strength $\gamma_\mathrm{PN}$. As indicated by the lensing potential derivation, a theoretical model with a stronger deviation from GR (larger $|\gamma_\mathrm{PN} - 1|$) can mimic the observational signature of a weaker deviation acting over a longer optical path (smaller $\Lambda$). Furthermore, while the phenomenological step-function model adopted in this work serves as a generalized probe, realistic modified gravity theories offer mechanisms to further break this degeneracy. In physical models such as the Chameleon mechanism or Vainshtein screening, the transition profile is not arbitrary but is dynamically coupled to the lens mass and environmental density. This imposes a rigid shape constraint on the potential, reducing the freedom to trade interaction strength for screening radius; however, using specific modified gravity models introduces a model dependency into the analysis.

For the final constraints on $\gamma_\mathrm{PN}$, using the power-law model in this paper as an example, the dominant sources of uncertainty are the measurement of magnification and the modeling of the lens galaxy. The magnification measurement error is fundamentally limited by ``lensing noise" -- scatter induced by weak lensing from large-scale structures along the line of sight -- which represents an irreducible astrophysical noise source. Regarding the lens modeling, it is anticipated that the precision of mass profile reconstruction can be enhanced with high-resolution data from future telescopes. However, significant systematic uncertainties still stem from the idealized assumptions in our model, namely that the lens galaxy's mass distribution follows a constant power-law and is perfectly spherically symmetric.
We will address these modeling limitations in the future work.

\section{Conclusions}

In this paper, we present a refined theoretical and statistical framework for probing screened modified gravity through strongly lensed gravitational waves. The approach is grounded in a rigorous Bayesian inference framework, which comprehensively incorporates observational uncertainties to enable precise quantification of degeneracies between lens mass profiles and modified gravity parameters. By introducing a mass-truncated power-law model, a critical mathematical divergence inherent in previous formulations near the isothermal limit ($\gamma' \approx 2$) is successfully corrected, thereby ensuring physical stability for realistic lens systems. Furthermore, the scope of the framework is expanded by providing an explicit derivation of the modified lensing potential for the NFW profile, yielding a robust and physically consistent tool for testing gravity across different dark matter halo structures.

The methodology emphasizes the synergistic potential between lensed gravitational wave events and collaborative observations of their host galaxies. It is highlighted that the inherent capability of gravitational waves to ascertain absolute magnification serves as the fundamental mechanism to break the MSD. To demonstrate the efficacy of this approach, a suite of lensed gravitational wave events was simulated, with the reconstructed posterior distributions for two representative cases presented to illustrate the robust recovery of lens and gravity parameters. Through this approach, we also quantified the projected constraints on the post-Newtonian parameter $\gamma_\mathrm{PN}$ across various screening scales $\Lambda$. The analysis indicates that systems with large Einstein radii and low source redshifts offer optimal sensitivity. Consequently, in the era of next-generation detectors like ET and DECIGO, this framework is expected to become a valuable tool for constraining the gravitational slip effect and testing the validity of GR.

Looking forward, the practical application of this framework will rely on electromagnetic observations that directly inform the ingredients already assumed in our lens modeling and Bayesian inference. In particular, high-resolution imaging of the lensed host galaxy, for instance with James Webb Space Telescope (JWST) \cite{2006SSRv..123..485G}, can constrain the lens mass profile and help control systematics associated with the adopted parametric models (e.g., the mass-truncated power-law or NFW profile). Wide-field optical imaging surveys such as Large Synoptic Survey Telescope (LSST) will be crucial for identifying and characterizing large samples of strong-lens candidates \cite{collettPOPULATIONGALAXYGALAXY2015}, providing valuable targets for follow-up observations. In addition, spatially resolved stellar kinematics of the lens galaxy can provide an independent handle on the mass normalization, thereby reducing degeneracies in the reconstruction. Moreover, combining the time delays between multiple lensed GW signals with the absolute magnification information accessible from lensed GW events enables a more robust breaking of the MSD and improves the joint constraints on $\gamma_\mathrm{PN}$ and the screening scale $\Lambda$ within the phenomenological screening framework adopted in this work. In parallel, next-generation GW detectors such as ET and DECIGO will increase the number of detectable lensed GW events and improve the measurement precision of time delays and lensing magnification, further enhancing the scientific return of joint EM+GW analyses.

\begin{acknowledgments}

This work is supported by Beijing Natural Science Foundation No. 1242021; the National Natural Science Foundation of China (Nos. 12021003, 12203009, 12433001); and the Strategic Priority Research Program of the Chinese Academy of Sciences, Grant No. XDB23000000.

\end{acknowledgments}

\bibliography{Screening}

@ARTICLE{2019NatSR...911608C,
       author = {{Cao}, Shuo and {Qi}, Jingzhao and {Cao}, Zhoujian and {Biesiada}, Marek and {Li}, Jin and {Pan}, Yu and {Zhu}, Zong-Hong},
        title = "{Direct test of the FLRW metric from strongly lensed gravitational wave observations}",
      journal = {Scientific Reports},
         year = 2019,
        month = aug,
       volume = {9},
          eid = {11608},
        pages = {11608},
          doi = {10.1038/s41598-019-47616-4},
archivePrefix = {arXiv},
       eprint = {1910.10365},
 primaryClass = {astro-ph.CO},
       adsurl = {https://ui.adsabs.harvard.edu/abs/2019NatSR...911608C}
}

@ARTICLE{2019PhRvD.100b3530Q,
       author = {{Qi}, Jingzhao and {Cao}, Shuo and {Biesiada}, Marek and {Zheng}, Xiaogang and {Ding}, Xuheng and {Zhu}, Zong-Hong},
        title = "{Strongly gravitationally lensed type Ia supernovae: Direct test of the Friedman-Lema{\^\i}tre-Robertson-Walker metric}",
      journal = {\prd},
         year = 2019,
        month = jul,
       volume = {100},
       number = {2},
          eid = {023530},
        pages = {023530},
          doi = {10.1103/PhysRevD.100.023530},
archivePrefix = {arXiv},
       eprint = {1802.05532},
 primaryClass = {astro-ph.CO},
       adsurl = {https://ui.adsabs.harvard.edu/abs/2019PhRvD.100b3530Q}
}

@ARTICLE{2016MNRAS.461.2192C,
       author = {{Cao}, Shuo and {Biesiada}, Marek and {Yao}, Meng and {Zhu}, Zong-Hong},
        title = "{Limits on the power-law mass and luminosity density profiles of elliptical galaxies from gravitational lensing systems}",
      journal = {\mnras},
         year = 2016,
        month = sep,
       volume = {461},
       number = {2},
        pages = {2192-2199},
          doi = {10.1093/mnras/stw932},
archivePrefix = {arXiv},
       eprint = {1604.05625},
 primaryClass = {astro-ph.CO},
       adsurl = {https://ui.adsabs.harvard.edu/abs/2016MNRAS.461.2192C}
}

@ARTICLE{2015ApJ...806..185C,
       author = {{Cao}, Shuo and {Biesiada}, Marek and {Gavazzi}, Rapha{\"e}l and {Pi{\'o}rkowska}, Aleksandra and {Zhu}, Zong-Hong},
        title = "{Cosmology with Strong-lensing Systems}",
      journal = {\apj},
         year = 2015,
        month = jun,
       volume = {806},
       number = {2},
          eid = {185},
        pages = {185},
          doi = {10.1088/0004-637X/806/2/185},
archivePrefix = {arXiv},
       eprint = {1509.07649},
 primaryClass = {astro-ph.CO},
       adsurl = {https://ui.adsabs.harvard.edu/abs/2015ApJ...806..185C}
}

@ARTICLE{2021MNRAS.502L..16C,
       author = {{Cao}, Shuo and {Qi}, Jingzhao and {Biesiada}, Marek and {Liu}, Tonghua and {Li}, Jin and {Zhu}, Zong-Hong},
        title = "{Measuring the viscosity of dark matter with strongly lensed gravitational waves}",
      journal = {\mnras},
         year = 2021,
        month = mar,
       volume = {502},
       number = {1},
        pages = {L16-L20},
          doi = {10.1093/mnrasl/slaa205},
archivePrefix = {arXiv},
       eprint = {2012.12462},
 primaryClass = {astro-ph.CO},
       adsurl = {https://ui.adsabs.harvard.edu/abs/2021MNRAS.502L..16C}
}

@ARTICLE{2022A&A...659L...5C,
       author = {{Cao}, Shuo and {Qi}, Jingzhao and {Cao}, Zhoujian and {Biesiada}, Marek and {Cheng}, Wei and {Zhu}, Zong-Hong},
        title = "{Direct measurement of the distribution of dark matter with strongly lensed gravitational waves}",
      journal = {\aap},
         year = 2022,
        month = mar,
       volume = {659},
          eid = {L5},
        pages = {L5},
          doi = {10.1051/0004-6361/202142694},
archivePrefix = {arXiv},
       eprint = {2202.08714},
 primaryClass = {astro-ph.CO},
       adsurl = {https://ui.adsabs.harvard.edu/abs/2022A&A...659L...5C}
}

@ARTICLE{2022ApJ...941...16L,
       author = {{Lian}, Yujie and {Cao}, Shuo and {Liu}, Tonghua and {Biesiada}, Marek and {Zhu}, Zong-Hong},
        title = "{Direct Tests of General Relativity under Screening Effect with Galaxy-scale Strong Lensing Systems}",
      journal = {\apj},
         year = 2022,
        month = dec,
       volume = {941},
       number = {1},
          eid = {16},
        pages = {16},
          doi = {10.3847/1538-4357/ac9d36},
archivePrefix = {arXiv},
       eprint = {2210.16752},
 primaryClass = {astro-ph.CO},
       adsurl = {https://ui.adsabs.harvard.edu/abs/2022ApJ...941...16L}
}

@ARTICLE{2017ApJ...835...92C,
       author = {{Cao}, Shuo and {Li}, Xiaolei and {Biesiada}, Marek and {Xu}, Tengpeng and {Cai}, Yongzhi and {Zhu}, Zong-Hong},
        title = "{Test of Parameterized Post-Newtonian Gravity with Galaxy-scale Strong Lensing Systems}",
      journal = {\apj},
         year = 2017,
        month = jan,
       volume = {835},
       number = {1},
          eid = {92},
        pages = {92},
          doi = {10.3847/1538-4357/835/1/92},
archivePrefix = {arXiv},
       eprint = {1701.00357},
 primaryClass = {astro-ph.CO},
       adsurl = {https://ui.adsabs.harvard.edu/abs/2017ApJ...835...92C}
}

@ARTICLE{2020ApJ...888L..25C,
       author = {{Cao}, Shuo and {Qi}, Jingzhao and {Biesiada}, Marek and {Liu}, Tonghua and {Zhu}, Zong-Hong},
        title = "{Precise Measurements of the Speed of Light with High-redshift Quasars: Ultra-compact Radio Structure and Strong Gravitational Lensing}",
      journal = {\apjl},
         year = 2020,
        month = jan,
       volume = {888},
       number = {2},
          eid = {L25},
        pages = {L25},
          doi = {10.3847/2041-8213/ab63d6},
       adsurl = {https://ui.adsabs.harvard.edu/abs/2020ApJ...888L..25C}
}

@article{abbottProspectsObservingLocalizing2020,
  title = {Prospects for Observing and Localizing Gravitational-Wave Transients with {{Advanced LIGO}}, {{Advanced Virgo}} and {{KAGRA}}},
  author = {{LIGO Scientific Collaboration} and {Virgo Collaboration} and {KAGRA Collaboration}},
  year = 2020,
  month = sep,
  journal = {Living Reviews in Relativity},
  volume = {23},
  number = {1},
  pages = {3},
  issn = {1433-8351},
  doi = {10.1007/s41114-020-00026-9},
  urldate = {2026-01-28},
  langid = {english}
}

@ARTICLE{2006SSRv..123..485G,
       author = {{Gies}, Holger},
        title = "{Astrophysical tests of quantum electrodynamics}",
      journal = {Space Science Reviews},
         year = 2006,
        month = nov,
       volume = {123},
       number = {1-3},
        pages = {485--505},
          doi = {10.1007/s11214-006-9088-5}
}

@article{bakerNovelProbesProject2021,
  title = {Novel {{Probes Project}}: {{Tests}} of Gravity on Astrophysical Scales},
  shorttitle = {Novel {{Probes Project}}},
  author = {Baker, Tessa and Barreira, Alexandre and Desmond, Harry and Ferreira, Pedro and Jain, Bhuvnesh and Koyama, Kazuya and Li, Baojiu and Lombriser, Lucas and Nicola, Andrina and Sakstein, Jeremy and Schmidt, Fabian},
  year = 2021,
  month = feb,
  journal = {Reviews of Modern Physics},
  volume = {93},
  number = {1},
  pages = {015003},
  issn = {0034-6861, 1539-0756},
  doi = {10.1103/RevModPhys.93.015003},
  urldate = {2025-03-04},
  langid = {english}
}

@article{biesiadaStrongGravitationalLensing2014,
  title = {Strong Gravitational Lensing of Gravitational Waves from Double Compact Binaries---Perspectives for the {{Einstein Telescope}}},
  author = {Biesiada, Marek and Ding, Xuheng and Pi{\'o}rkowska, Aleksandra and Zhu, Zong-Hong},
  year = 2014,
  month = oct,
  journal = {Journal of Cosmology and Astroparticle Physics},
  volume = {2014},
  number = {10},
  pages = {080--080},
  issn = {1475-7516},
  doi = {10.1088/1475-7516/2014/10/080},
  urldate = {2024-11-18},
  copyright = {http://iopscience.iop.org/info/page/text-and-data-mining},
  langid = {english}
}

@article{carrollCosmologicalConstant2001,
  title = {The {{Cosmological Constant}}},
  author = {Carroll, Sean M.},
  year = 2001,
  month = feb,
  journal = {Living Reviews in Relativity},
  volume = {4},
  number = {1},
  pages = {1},
  issn = {1433-8351},
  doi = {10.12942/lrr-2001-1},
  urldate = {2025-12-17},
  langid = {english}
}

@article{collaborationGravitationalWavesGammarays2017,
  title = {Gravitational {{Waves}} and {{Gamma-rays}} from a {{Binary Neutron Star Merger}}: {{GW170817}} and {{GRB 170817A}}},
  shorttitle = {Gravitational {{Waves}} and {{Gamma-rays}} from a {{Binary Neutron Star Merger}}},
  author = {{LIGO Scientific Collaboration} and {Virgo Collaboration} and {Fermi Gamma-Ray Burst Monitor} and {INTEGRAL}},
  year = 2017,
  month = oct,
  journal = {The Astrophysical Journal Letters},
  volume = {848},
  number = {2},
  eprint = {1710.05834},
  primaryclass = {astro-ph},
  pages = {L13},
  issn = {2041-8205, 2041-8213},
  doi = {10.3847/2041-8213/aa920c},
  urldate = {2026-01-28},
  archiveprefix = {arXiv}
}

@article{collaborationGW150914FirstResults2016,
  title = {{{GW150914}}: {{First}} Results from the Search for Binary Black Hole Coalescence with {{Advanced LIGO}}},
  shorttitle = {{{GW150914}}},
  author = {{LIGO Scientific Collaboration} and {Virgo Collaboration}},
  year = 2016,
  month = jun,
  journal = {Physical Review D},
  volume = {93},
  number = {12},
  eprint = {1602.03839},
  primaryclass = {gr-qc},
  pages = {122003},
  issn = {2470-0010, 2470-0029},
  doi = {10.1103/PhysRevD.93.122003},
  urldate = {2026-01-28},
  archiveprefix = {arXiv}
}

@article{collaborationGW170817ObservationGravitational2017,
  title = {{{GW170817}}: {{Observation}} of {{Gravitational Waves}} from a {{Binary Neutron Star Inspiral}}},
  shorttitle = {{{GW170817}}},
  author = {{LIGO Scientific Collaboration} and {Virgo Collaboration}},
  year = 2017,
  month = oct,
  journal = {Physical Review Letters},
  volume = {119},
  number = {16},
  eprint = {1710.05832},
  primaryclass = {gr-qc},
  pages = {161101},
  issn = {0031-9007, 1079-7114},
  doi = {10.1103/PhysRevLett.119.161101},
  urldate = {2025-12-14},
  archiveprefix = {arXiv}
}

@article{collaborationPlanck2018Results2020,
  title = {Planck 2018 Results. {{I}}. {{Overview}} and the Cosmological Legacy of {{Planck}}},
  author = {{Planck Collaboration}},
  year = 2020,
  month = sep,
  journal = {Astronomy \& Astrophysics},
  volume = {641},
  eprint = {1807.06205},
  primaryclass = {astro-ph},
  pages = {A1},
  issn = {0004-6361, 1432-0746},
  doi = {10.1051/0004-6361/201833880},
  urldate = {2026-01-28},
  archiveprefix = {arXiv}
}

@article{collettPOPULATIONGALAXYGALAXY2015,
author = {{Collett}, Thomas E.},
        title = "{The Population of Galaxy-Galaxy Strong Lenses in Forthcoming Optical Imaging Surveys}",
      journal = {\apj},
         year = 2015,
        month = sep,
       volume = {811},
       number = {1},
          eid = {20},
        pages = {20},
          doi = {10.1088/0004-637X/811/1/20},
archivePrefix = {arXiv},
       eprint = {1507.02657},
 primaryClass = {astro-ph.CO},
       adsurl = {https://ui.adsabs.harvard.edu/abs/2015ApJ...811...20C}
}

@article{collettPreciseExtragalacticTest2018,
  title = {A Precise Extragalactic Test of {{General Relativity}}},
  author = {Collett, Thomas E. and Oldham, Lindsay J. and Smith, Russell J. and Auger, Matthew W. and Westfall, Kyle B. and Bacon, David and Nichol, Robert C. and Masters, Karen L. and Koyama, Kazuya and van den Bosch, Remco},
  year = 2018,
  month = jun,
  journal = {Science},
  volume = {360},
  number = {6395},
  eprint = {1806.08300},
  primaryclass = {astro-ph},
  pages = {1342--1346},
  issn = {0036-8075, 1095-9203},
  doi = {10.1126/science.aao2469},
  urldate = {2026-01-28},
  archiveprefix = {arXiv}
}

@article{deffayetNonperturbativeContinuityGraviton2002,
  title = {Nonperturbative Continuity in Graviton Mass versus Perturbative Discontinuity},
  author = {Deffayet, C{\'e}dric and Dvali, Gia and Gabadadze, Gregory and Vainshtein, Arkady},
  year = 2002,
  month = jan,
  journal = {Physical Review D},
  volume = {65},
  number = {4},
  pages = {044026},
  publisher = {American Physical Society},
  doi = {10.1103/PhysRevD.65.044026},
  urldate = {2026-01-28}
}

@article{foreman-mackeyEmceeMCMCHammer2013,
  title = {Emcee: {{The MCMC Hammer}}},
  shorttitle = {Emcee},
  author = {{Foreman-Mackey}, Daniel and Hogg, David W. and Lang, Dustin and Goodman, Jonathan},
  year = 2013,
  month = mar,
  journal = {Publications of the Astronomical Society of the Pacific},
  volume = {125},
  number = {925},
  eprint = {1202.3665},
  primaryclass = {astro-ph},
  pages = {306--312},
  issn = {00046280, 15383873},
  doi = {10.1086/670067},
  urldate = {2026-01-30},
  archiveprefix = {arXiv}
}

@article{gunnInfallMatterClusters1972,
  title = {On the {{Infall}} of {{Matter Into Clusters}} of {{Galaxies}} and {{Some Effects}} on {{Their Evolution}}},
  author = {Gunn, James E. and Gott, Iii, J. Richard},
  year = 1972,
  month = aug,
  journal = {The Astrophysical Journal},
  volume = {176},
  pages = {1},
  issn = {0004-637X, 1538-4357},
  doi = {10.1086/151605},
  urldate = {2026-01-30},
  langid = {english}
}

@misc{hannukselaLocalizingMergingBlack2020,
  title = {Localizing Merging Black Holes with Sub-Arcsecond Precision Using Gravitational-Wave Lensing},
  author = {Hannuksela, Otto A. and Collett, Thomas E. and {\c C}al{\i}{\c s}kan, Mesut and Li, Tjonnie G. F.},
  year = 2020,
  month = aug,
  eprint = {2004.13811},
  primaryclass = {astro-ph},
  doi = {10.1093/mnras/staa2577},
  urldate = {2026-01-17},
  archiveprefix = {arXiv},
  langid = {english}
}

@article{hinterbichlerScreeningLongRangeForces2010,
  title = {Screening {{Long-Range Forces}} through {{Local Symmetry Restoration}}},
  author = {Hinterbichler, Kurt and Khoury, Justin},
  year = 2010,
  month = jun,
  journal = {Physical Review Letters},
  volume = {104},
  number = {23},
  pages = {231301},
  publisher = {American Physical Society},
  doi = {10.1103/PhysRevLett.104.231301},
  urldate = {2026-01-28}
}

@article{holzUsingGravitationalwaveStandard2005,
  title = {Using Gravitational-Wave Standard Sirens},
  author = {Holz, Daniel E. and Hughes, Scott A.},
  year = 2005,
  month = aug,
  journal = {The Astrophysical Journal},
  volume = {629},
  number = {1},
  eprint = {astro-ph/0504616},
  pages = {15--22},
  issn = {0004-637X, 1538-4357},
  doi = {10.1086/431341},
  urldate = {2026-01-29},
  archiveprefix = {arXiv}
}

@article{huModelsCosmicAcceleration2007,
  title = {Models of f ( {{R}} ) Cosmic Acceleration That Evade Solar System Tests},
  author = {Hu, Wayne and Sawicki, Ignacy},
  year = 2007,
  month = sep,
  journal = {Physical Review D},
  volume = {76},
  number = {6},
  pages = {064004},
  issn = {1550-7998, 1550-2368},
  doi = {10.1103/PhysRevD.76.064004},
  urldate = {2025-04-04},
  copyright = {http://link.aps.org/licenses/aps-default-license},
  langid = {english}
}

@article{isoyamaMultibandGravitationalWaveAstronomy2018,
  title = {Multiband {{Gravitational-Wave Astronomy}}: {{Observing}} Binary Inspirals with a Decihertz Detector, {{B-DECIGO}}},
  shorttitle = {Multiband {{Gravitational-Wave Astronomy}}},
  author = {Isoyama, Soichiro and Nakano, Hiroyuki and Nakamura, Takashi},
  year = 2018,
  month = jul,
  journal = {Progress of Theoretical and Experimental Physics},
  volume = {2018},
  number = {7},
  eprint = {1802.06977},
  primaryclass = {gr-qc},
  issn = {2050-3911},
  doi = {10.1093/ptep/pty078},
  urldate = {2025-01-01},
  archiveprefix = {arXiv},
  langid = {english}
}

@article{jyotiCosmicTimeSlip2019,
  title = {Cosmic Time Slip: {{Testing}} Gravity on Supergalactic Scales with Strong-Lensing Time Delays},
  shorttitle = {Cosmic Time Slip},
  author = {Jyoti, Dhrubo and Mu{\~n}oz, Julian B. and Caldwell, Robert R. and Kamionkowski, Marc},
  year = 2019,
  month = aug,
  journal = {Physical Review D},
  volume = {100},
  number = {4},
  pages = {043031},
  issn = {2470-0010, 2470-0029},
  doi = {10.1103/PhysRevD.100.043031},
  urldate = {2026-01-28},
  langid = {english}
}

@misc{kawamuraCurrentStatusSpace2020,
  title = {Current Status of Space Gravitational Wave Antenna {{DECIGO}} and {{B-DECIGO}}},
  author = {Kawamura, Seiji and Ando, Masaki and Seto, Naoki and Sato, Shuichi and Musha, Mitsuru and Kawano, Isao and Yokoyama, Jun'ichi and Tanaka, Takahiro and Ioka, Kunihito and Akutsu, Tomotada and Takashima, Takeshi and Agatsuma, Kazuhiro and Araya, Akito and Aritomi, Naoki and Asada, Hideki and Chiba, Takeshi and Eguchi, Satoshi and Enoki, Motohiro and Fujimoto, Masa-Katsu and Fujita, Ryuichi and Futamase, Toshifumi and Harada, Tomohiro and Hayama, Kazuhiro and Himemoto, Yoshiaki and Hiramatsu, Takashi and Hong, Feng-Lei and Hosokawa, Mizuhiko and Ichiki, Kiyotomo and Ikari, Satoshi and Ishihara, Hideki and Ishikawa, Tomohiro and Itoh, Yousuke and Ito, Takahiro and Iwaguchi, Shoki and Izumi, Kiwamu and Kanda, Nobuyuki and Kanemura, Shinya and Kawazoe, Fumiko and Kobayashi, Shiho and Kohri, Kazunori and Kojima, Yasufumi and Kokeyama, Keiko and Kotake, Kei and Kuroyanagi, Sachiko and Maeda, Kei-ichi and Matsushita, Shuhei and Michimura, Yuta and Morimoto, Taigen and Mukohyama, Shinji and Nagano, Koji and Nagano, Shigeo and Naito, Takeo and Nakamura, Kouji and Nakamura, Takashi and Nakano, Hiroyuki and Nakao, Kenichi and Nakasuka, Shinichi and Nakayama, Yoshinori and Nakazawa, Kazuhiro and Nishizawa, Atsushi and Ohkawa, Masashi and Oohara, Kenichi and Sago, Norichika and Saijo, Motoyuki and Sakagami, Masaaki and Sakai, Shin-ichiro and Sato, Takashi and Shibata, Masaru and Shinkai, Hisaaki and Shoda, Ayaka and Somiya, Kentaro and Sotani, Hajime and Takahashi, Ryutaro and Takahashi, Hirotaka and Akiteru, Takamori and Taniguchi, Keisuke and Taruya, Atsushi and Tsubono, Kimio and Tsujikawa, Shinji and Ueda, Akitoshi and Ueda, Ken-ichi and Watanabe, Izumi and Yagi, Kent and Yamada, Rika and Yokoyama, Shuichiro and Yoo, Chul-Moon and Zhu, Zong-Hong},
  year = 2020,
  month = jun,
  number = {arXiv:2006.13545},
  eprint = {2006.13545},
  primaryclass = {gr-qc},
  publisher = {arXiv},
  doi = {10.48550/arXiv.2006.13545},
  urldate = {2026-01-28},
  archiveprefix = {arXiv}
}

@article{khouryChameleonCosmology2004,
  title = {Chameleon Cosmology},
  author = {Khoury, Justin and Weltman, Amanda},
  year = 2004,
  journal = {Phys. Rev. D},
  volume = {69},
  pages = {044026},
  doi = {10.1103/PhysRevD.69.044026}
}

@article{khouryChameleonFieldsAwaiting2004,
  title = {Chameleon {{Fields}}: {{Awaiting Surprises}} for {{Tests}} of {{Gravity}} in {{Space}}},
  shorttitle = {Chameleon {{Fields}}},
  author = {Khoury, Justin and Weltman, Amanda},
  year = 2004,
  month = oct,
  journal = {Physical Review Letters},
  volume = {93},
  number = {17},
  pages = {171104},
  publisher = {American Physical Society},
  doi = {10.1103/PhysRevLett.93.171104},
  urldate = {2026-01-28}
}

@article{koopmansGravitationalLensingStellar2006,
  title = {Gravitational {{Lensing}} \& {{Stellar Dynamics}}},
  author = {Koopmans, L. V. E.},
  year = 2006,
  journal = {EAS Publications Series},
  volume = {20},
  eprint = {astro-ph/0511121},
  pages = {161--166},
  issn = {1633-4760, 1638-1963},
  doi = {10.1051/eas:2006064},
  urldate = {2025-05-12},
  archiveprefix = {arXiv},
  langid = {english}
}

@book{maggioreGravitationalWaves2008,
  title = {Gravitational Waves},
  author = {Maggiore, Michele},
  year = 2008,
  publisher = {Oxford University Press},
  address = {Oxford},
  isbn = {978-0-19-857074-5 978-0-19-857089-9},
  langid = {english},
  lccn = {QC179 .M34 2008}
}

@article{nairSynergyGroundSpace2018,
  title = {Synergy between Ground and Space Based Gravitational Wave Detectors. {{Part II}}: {{Localisation}}},
  shorttitle = {Synergy between Ground and Space Based Gravitational Wave Detectors. {{Part II}}},
  author = {Nair, Remya and Tanaka, Takahiro},
  year = 2018,
  month = aug,
  journal = {Journal of Cosmology and Astroparticle Physics},
  volume = {2018},
  number = {08},
  pages = {033--033},
  issn = {1475-7516},
  doi = {10.1088/1475-7516/2018/08/033},
  urldate = {2025-01-05},
  copyright = {http://iopscience.iop.org/info/page/text-and-data-mining},
  langid = {english}
}

@misc{narayanLecturesGravitationalLensing1997,
  title = {Lectures on {{Gravitational Lensing}}},
  author = {Narayan, Ramesh and Bartelmann, Matthias},
  year = 1997,
  month = oct,
  number = {arXiv:astro-ph/9606001},
  eprint = {astro-ph/9606001},
  publisher = {arXiv},
  urldate = {2024-01-01},
  archiveprefix = {arXiv},
  langid = {english}
}

@article{navarroUniversalDensityProfile1997,
  title = {A {{Universal Density Profile}} from {{Hierarchical Clustering}}},
  author = {Navarro, Julio F. and Frenk, Carlos S. and White, Simon D. M.},
  year = 1997,
  month = dec,
  journal = {The Astrophysical Journal},
  volume = {490},
  number = {2},
  pages = {493},
  publisher = {IOP Publishing},
  issn = {0004-637X},
  doi = {10.1086/304888},
  urldate = {2026-01-28},
  langid = {english}
}

@article{nicolisGalileonLocalModification2009,
  title = {Galileon as a Local Modification of Gravity},
  author = {Nicolis, Alberto and Rattazzi, Riccardo and Trincherini, Enrico},
  year = 2009,
  month = mar,
  journal = {Physical Review D},
  volume = {79},
  number = {6},
  pages = {064036},
  publisher = {American Physical Society},
  doi = {10.1103/PhysRevD.79.064036},
  urldate = {2026-01-28}
}

@article{peeblesCosmologicalConstantDark2003,
  title = {The {{Cosmological Constant}} and {{Dark Energy}}},
  author = {Peebles, P. J. E. and Ratra, Bharat},
  year = 2003,
  month = apr,
  journal = {Reviews of Modern Physics},
  volume = {75},
  number = {2},
  eprint = {astro-ph/0207347},
  pages = {559--606},
  issn = {0034-6861, 1539-0756},
  doi = {10.1103/RevModPhys.75.559},
  urldate = {2026-01-28},
  archiveprefix = {arXiv},
  langid = {english}
}

@article{perlmutterMeasurements42HighRedshift1999,
  title = {Measurements of {{\textohm}} and {{$\Lambda$}} from 42 {{High-Redshift Supernovae}}},
  author = {Perlmutter, S. and Aldering, G. and Goldhaber, G. and Knop, R. A. and Nugent, P. and Castro, P. G. and Deustua, S. and Fabbro, S. and Goobar, A. and Groom, D. E. and Hook, I. M. and Kim, A. G. and Kim, M. Y. and Lee, J. C. and Nunes, N. J. and Pain, R. and Pennypacker, C. R. and Quimby, R. and Lidman, C. and Ellis, R. S. and Irwin, M. and McMahon, R. G. and {Ruiz-Lapuente}, P. and Walton, N. and Schaefer, B. and Boyle, B. J. and Filippenko, A. V. and Matheson, T. and Fruchter, A. S. and Panagia, N. and Newberg, H. J. M. and Couch, W. J. and Project, The Supernova Cosmology},
  year = 1999,
  month = jun,
  journal = {The Astrophysical Journal},
  volume = {517},
  number = {2},
  pages = {565},
  publisher = {IOP Publishing},
  issn = {0004-637X},
  doi = {10.1086/307221},
  urldate = {2026-01-26},
  langid = {english}
}

@article{piorkowska-kurpasInspiralingDoubleCompact2021,
  title = {Inspiraling Double Compact Object Detection and Lensing Rate -- Forecast for {{DECIGO}} and {{B-DECIGO}}},
  author = {{Pi{\'o}rkowska-Kurpas}, Aleksandra and Hou, Shaoqi and Biesiada, Marek and Ding, Xuheng and Cao, Shuo and Fan, Xilong and Kawamura, Seiji and Zhu, Zong-Hong},
  year = 2021,
  month = feb,
  journal = {The Astrophysical Journal},
  volume = {908},
  number = {2},
  eprint = {2005.08727},
  primaryclass = {astro-ph},
  pages = {196},
  issn = {0004-637X, 1538-4357},
  doi = {10.3847/1538-4357/abd482},
  urldate = {2024-11-16},
  archiveprefix = {arXiv},
  langid = {english}
}

@misc{piorkowskaStrongGravitationalLensing2013,
  title = {Strong Gravitational Lensing of Gravitational Waves in {{Einstein Telescope}}},
  author = {Pi{\'o}rkowska, Aleksandra and Biesiada, Marek and Zhu, Zong-Hong},
  year = 2013,
  month = sep,
  eprint = {1309.5731},
  primaryclass = {astro-ph},
  doi = {10.1088/1475-7516/2013/10/022},
  urldate = {2024-11-18},
  archiveprefix = {arXiv},
  langid = {english}
}

@article{punturoEinsteinTelescopeThirdgeneration2010,
  title = {The {{Einstein Telescope}}: A Third-Generation Gravitational Wave Observatory},
  shorttitle = {The {{Einstein Telescope}}},
  author = {Punturo, M and Abernathy, M and Acernese, F and Allen, B and Andersson, N and Arun, K and Barone, F and Barr, B and Barsuglia, M and Beker, M and Beveridge, N and Birindelli, S and Bose, S and Bosi, L and Braccini, S and Bradaschia, C and Bulik, T and Calloni, E and Cella, G and Mottin, E Chassande and Chelkowski, S and Chincarini, A and Clark, J and Coccia, E and Colacino, C and Colas, J and Cumming, A and Cunningham, L and Cuoco, E and Danilishin, S and Danzmann, K and De Luca, G and De Salvo, R and Dent, T and De Rosa, R and Di Fiore, L and Di Virgilio, A and Doets, M and Fafone, V and Falferi, P and Flaminio, R and Franc, J and Frasconi, F and Freise, A and Fulda, P and Gair, J and Gemme, G and Gennai, A and Giazotto, A and Glampedakis, K and Granata, M and Grote, H and Guidi, G and Hammond, G and Hannam, M and Harms, J and Heinert, D and Hendry, M and Heng, I and Hennes, E and Hild, S and Hough, J and Husa, S and Huttner, S and Jones, G and Khalili, F and Kokeyama, K and Kokkotas, K and Krishnan, B and Lorenzini, M and L{\"u}ck, H and Majorana, E and Mandel, I and Mandic, V and Martin, I and Michel, C and Minenkov, Y and Morgado, N and Mosca, S and Mours, B and {M{\"u}ller--Ebhardt}, H and Murray, P and Nawrodt, R and Nelson, J and Oshaughnessy, R and Ott, C D and Palomba, C and Paoli, A and Parguez, G and Pasqualetti, A and Passaquieti, R and Passuello, D and Pinard, L and Poggiani, R and Popolizio, P and Prato, M and Puppo, P and Rabeling, D and Rapagnani, P and Read, J and Regimbau, T and Rehbein, H and Reid, S and Rezzolla, L and Ricci, F and Richard, F and Rocchi, A and Rowan, S and R{\"u}diger, A and Sassolas, B and Sathyaprakash, B and Schnabel, R and Schwarz, C and Seidel, P and Sintes, A and Somiya, K and Speirits, F and Strain, K and Strigin, S and Sutton, P and Tarabrin, S and Th{\"u}ring, A and {van den Brand}, J and {van Leewen}, C and {van Veggel}, M and {van den Broeck}, C and Vecchio, A and Veitch, J and Vetrano, F and Vicere, A and Vyatchanin, S and Willke, B and Woan, G and Wolfango, P and Yamamoto, K},
  year = 2010,
  month = sep,
  journal = {Classical and Quantum Gravity},
  volume = {27},
  number = {19},
  pages = {194002},
  issn = {0264-9381},
  doi = {10.1088/0264-9381/27/19/194002},
  urldate = {2026-01-28},
  langid = {english}
}

@article{riessExpansionUniverseFaster2019,
  title = {The {{Expansion}} of the {{Universe}} Is {{Faster}} than {{Expected}}},
  author = {Riess, Adam G.},
  year = 2019,
  month = dec,
  journal = {Nature Reviews Physics},
  volume = {2},
  number = {1},
  eprint = {2001.03624},
  primaryclass = {astro-ph},
  pages = {10--12},
  issn = {2522-5820},
  doi = {10.1038/s42254-019-0137-0},
  urldate = {2026-01-28},
  archiveprefix = {arXiv}
}

@article{riessObservationalEvidenceSupernovae1998,
  title = {Observational {{Evidence}} from {{Supernovae}} for an {{Accelerating Universe}} and a {{Cosmological Constant}}},
  author = {Riess, Adam G. and Filippenko, Alexei V. and Challis, Peter and Clocchiatti, Alejandro and Diercks, Alan and Garnavich, Peter M. and Gilliland, Ron L. and Hogan, Craig J. and Jha, Saurabh and Kirshner, Robert P. and Leibundgut, B. and Phillips, M. M. and Reiss, David and Schmidt, Brian P. and Schommer, Robert A. and Smith, R. Chris and Spyromilio, J. and Stubbs, Christopher and Suntzeff, Nicholas B. and Tonry, John},
  year = 1998,
  month = sep,
  journal = {The Astronomical Journal},
  volume = {116},
  number = {3},
  pages = {1009},
  publisher = {IOP Publishing},
  issn = {1538-3881},
  doi = {10.1086/300499},
  urldate = {2026-01-26},
  langid = {english}
}

@article{rusuH0LiCOWXIILens2020,
  title = {{{H0LiCOW XII}}. {{Lens}} Mass Model of {{WFI2033-4723}} and Blind Measurement of Its Time-Delay Distance and \${{H}}\_0\$},
  author = {Rusu, Cristian E. and Wong, Kenneth C. and Bonvin, Vivien and Sluse, Dominique and Suyu, Sherry H. and Fassnacht, Christopher D. and Chan, James H. H. and Hilbert, Stefan and Auger, Matthew W. and Sonnenfeld, Alessandro and Birrer, Simon and Courbin, Frederic and Treu, Tommaso and Chen, Geoff C.-F. and Halkola, Aleksi and Koopmans, Leon V. E. and Marshall, Philip J. and Shajib, Anowar J.},
  year = 2020,
  month = oct,
  journal = {Monthly Notices of the Royal Astronomical Society},
  volume = {498},
  number = {1},
  eprint = {1905.09338},
  primaryclass = {astro-ph},
  pages = {1440--1468},
  issn = {0035-8711, 1365-2966},
  doi = {10.1093/mnras/stz3451},
  urldate = {2026-01-29},
  archiveprefix = {arXiv}
}

@book{schneiderGravitationalLenses1992,
  title = {Gravitational {{Lenses}}},
  author = {Schneider, Peter and Ehlers, J{\"u}rgen and Falco, Emilio E.},
  editor = {Appenzeller, I. and B{\"o}rner, G. and Harwit, M. and Kippenhahn, R. and Lequeux, J. and Strittmatter, P. A. and Trimble, V.},
  year = 1992,
  series = {Astronomy and {{Astrophysics Library}}},
  publisher = {Springer Berlin Heidelberg},
  address = {Berlin, Heidelberg},
  doi = {10.1007/978-3-662-03758-4},
  urldate = {2024-01-01},
  isbn = {978-3-540-66506-9 978-3-662-03758-4},
  langid = {english}
}

@book{schneiderGravitationalLensingStrong2006,
  title = {Gravitational Lensing: Strong, Weak and Micro},
  shorttitle = {Gravitational Lensing},
  author = {Schneider, Peter and Kochanek, Christopher S. and Wambsganss, Joachim},
  year = 2006,
  publisher = {Springer},
  address = {Berlin New York},
  collaborator = {{Saas-fee advanced course}},
  isbn = {978-3-540-30309-1},
  langid = {english},
  lccn = {521.1}
}

@article{schneiderMasssheetDegeneracyPowerlaw2013,
  title = {Mass-Sheet Degeneracy, Power-Law Models and External Convergence: {{Impact}} on the Determination of the {{Hubble}} Constant from Gravitational Lensing},
  shorttitle = {Mass-Sheet Degeneracy, Power-Law Models and External Convergence},
  author = {Schneider, Peter and Sluse, Dominique},
  year = 2013,
  month = nov,
  journal = {Astronomy \& Astrophysics},
  volume = {559},
  eprint = {1306.0901},
  primaryclass = {astro-ph},
  pages = {A37},
  issn = {0004-6361, 1432-0746},
  doi = {10.1051/0004-6361/201321882},
  urldate = {2024-10-15},
  archiveprefix = {arXiv},
  langid = {english}
}

@article{schneiderSourcepositionTransformationApproximate2014,
  title = {Source-Position Transformation -- an Approximate Invariance in Strong Gravitational Lensing},
  author = {Schneider, Peter and Sluse, Dominique},
  year = 2014,
  month = apr,
  journal = {Astronomy \& Astrophysics},
  volume = {564},
  eprint = {1306.4675},
  primaryclass = {astro-ph},
  pages = {A103},
  issn = {0004-6361, 1432-0746},
  doi = {10.1051/0004-6361/201322106},
  urldate = {2024-11-30},
  archiveprefix = {arXiv},
  langid = {english}
}

@article{suyuH0LiCOW$H_0$Lenses2017,
  title = {{{H0LiCOW I}}. \${{H}}\_0\$ {{Lenses}} in {{COSMOGRAIL}}'s {{Wellspring}}: {{Program Overview}}},
  shorttitle = {{{H0LiCOW I}}. \${{H}}\_0\$ {{Lenses}} in {{COSMOGRAIL}}'s {{Wellspring}}},
  author = {Suyu, S. H. and Bonvin, V. and Courbin, F. and Fassnacht, C. D. and Rusu, C. E. and Sluse, D. and Treu, T. and Wong, K. C. and Auger, M. W. and Ding, X. and Hilbert, S. and Marshall, P. J. and Rumbaugh, N. and Sonnenfeld, A. and Tewes, M. and Tihhonova, O. and Agnello, A. and Blandford, R. D. and Chen, G. C.-F. and Collett, T. and Koopmans, L. V. E. and Liao, K. and Meylan, G. and Spiniello, C.},
  year = 2017,
  month = jul,
  journal = {Monthly Notices of the Royal Astronomical Society},
  volume = {468},
  number = {3},
  eprint = {1607.00017},
  primaryclass = {astro-ph},
  pages = {2590--2604},
  issn = {0035-8711, 1365-2966},
  doi = {10.1093/mnras/stx483},
  urldate = {2026-01-29},
  archiveprefix = {arXiv}
}

@article{suyuTwoAccurateTimedelay2013,
  title = {Two Accurate Time-Delay Distances from Strong Lensing: {{Implications}} for Cosmology},
  shorttitle = {Two Accurate Time-Delay Distances from Strong Lensing},
  author = {Suyu, S. H. and Auger, M. W. and Hilbert, S. and Marshall, P. J. and Tewes, M. and Treu, T. and Fassnacht, C. D. and Koopmans, L. V. E. and Sluse, D. and Blandford, R. D. and Courbin, F. and Meylan, G.},
  year = 2013,
  month = mar,
  journal = {The Astrophysical Journal},
  volume = {766},
  number = {2},
  eprint = {1208.6010},
  primaryclass = {astro-ph},
  pages = {70},
  issn = {0004-637X, 1538-4357},
  doi = {10.1088/0004-637X/766/2/70},
  urldate = {2024-11-21},
  archiveprefix = {arXiv},
  langid = {english}
}

@article{valentinoCosmologyIntertwinedII2021,
  title = {Cosmology {{Intertwined II}}: {{The Hubble Constant Tension}}},
  shorttitle = {Cosmology {{Intertwined II}}},
  author = {Valentino, Eleonora Di and Anchordoqui, Luis A. and Akarsu, Ozgur and {Ali-Haimoud}, Yacine and Amendola, Luca and Arendse, Nikki and Asgari, Marika and Ballardini, Mario and Basilakos, Spyros and Battistelli, Elia and Benetti, Micol and Birrer, Simon and Bouchet, Fran{\c c}ois R. and Bruni, Marco and Calabrese, Erminia and Camarena, David and Capozziello, Salvatore and Chen, Angela and Chluba, Jens and Chudaykin, Anton and Colg{\'a}in, Eoin {\'O} and {Cyr-Racine}, Francis-Yan and de Bernardis, Paolo and P{\'e}rez, Javier de Cruz and Delabrouille, Jacques and Dunkley, Jo and {Escamilla-Rivera}, Celia and Fert{\'e}, Agn{\`e}s and Finelli, Fabio and Freedman, Wendy and Frusciante, Noemi and Giusarma, Elena and {G{\'o}mez-Valent}, Adri{\`a} and Guy, Julien and Handley, Will and Harrison, Ian and Hart, Luke and Heavens, Alan and Hildebrandt, Hendrik and Holz, Daniel and Huterer, Dragan and Ivanov, Mikhail M. and Joudaki, Shahab and Kamionkowski, Marc and Karwal, Tanvi and Knox, Lloyd and Kumar, Suresh and Lamagna, Luca and Lesgourgues, Julien and Lucca, Matteo and Marra, Valerio and Masi, Silvia and Matarrese, Sabino and Mazumdar, Arindam and Melchiorri, Alessandro and Mena, Olga and {Mersini-Houghton}, Laura and Miranda, Vivian and {Moreno-Pulido}, Cristian and Mota, David F. and Muir, Jessica and Mukherjee, Ankan and Niedermann, Florian and Notari, Alessio and Nunes, Rafael C. and Pace, Francesco and Paliathanasis, Andronikos and Palmese, Antonella and Pan, Supriya and Paoletti, Daniela and Pettorino, Valeria and Piacentini, Francesco and Poulin, Vivian and Raveri, Marco and Riess, Adam G. and Salzano, Vincenzo and Saridakis, Emmanuel N. and Sen, Anjan A. and Shafieloo, Arman and Shajib, Anowar J. and Silk, Joseph and Silvestri, Alessandra and Sloth, Martin S. and Smith, Tristan L. and Sol{\`a}, Joan and van de Bruck, Carsten and Verde, Licia and Visinelli, Luca and Wandelt, Benjamin D. and Wang, Deng and Wang, Jian-Min and Yadav, Anil K. and Yang, Weiqiang},
  year = 2021,
  month = sep,
  journal = {Astroparticle Physics},
  volume = {131},
  eprint = {2008.11284},
  primaryclass = {astro-ph},
  pages = {102605},
  issn = {09276505},
  doi = {10.1016/j.astropartphys.2021.102605},
  urldate = {2026-01-28},
  archiveprefix = {arXiv}
}

@article{verdeTensionsEarlyLate2019,
  title = {Tensions between the {{Early}} and the {{Late Universe}}},
  author = {Verde, L. and Treu, T. and Riess, A. G.},
  year = 2019,
  month = sep,
  journal = {Nature Astronomy},
  volume = {3},
  number = {10},
  eprint = {1907.10625},
  primaryclass = {astro-ph},
  pages = {891--895},
  issn = {2397-3366},
  doi = {10.1038/s41550-019-0902-0},
  urldate = {2026-01-28},
  archiveprefix = {arXiv}
}

@article{willConfrontationGeneralRelativity2014,
  title = {The {{Confrontation}} between {{General Relativity}} and {{Experiment}}},
  author = {Will, Clifford M.},
  year = 2014,
  month = dec,
  journal = {Living Reviews in Relativity},
  volume = {17},
  number = {1},
  eprint = {1403.7377},
  primaryclass = {gr-qc},
  pages = {4},
  issn = {2367-3613, 1433-8351},
  doi = {10.12942/lrr-2014-4},
  urldate = {2026-01-28},
  archiveprefix = {arXiv}
}

@article{wongH0LiCOWIVLens2017,
  title = {{{H0LiCOW IV}}. {{Lens}} Mass Model of {{HE}} 0435-1223 and Blind Measurement of Its Time-Delay Distance for Cosmology},
  author = {Wong, Kenneth C. and Suyu, Sherry H. and Auger, Matthew W. and Bonvin, Vivien and Courbin, Frederic and Fassnacht, Christopher D. and Halkola, Aleksi and Rusu, Cristian E. and Sluse, Dominique and Sonnenfeld, Alessandro and Treu, Tommaso and Collett, Thomas E. and Hilbert, Stefan and Koopmans, Leon V. E. and Marshall, Philip J. and Rumbaugh, Nicholas},
  year = 2017,
  month = mar,
  journal = {Monthly Notices of the Royal Astronomical Society},
  volume = {465},
  number = {4},
  eprint = {1607.01403},
  primaryclass = {astro-ph},
  pages = {4895--4913},
  issn = {0035-8711, 1365-2966},
  doi = {10.1093/mnras/stw3077},
  urldate = {2026-01-29},
  archiveprefix = {arXiv}
}

@article{zhaoDeterminationDarkEnergy2011,
  title = {Determination of {{Dark Energy}} by the {{Einstein Telescope}}: {{Comparing}} with {{CMB}}, {{BAO}} and {{SNIa Observations}}},
  shorttitle = {Determination of {{Dark Energy}} by the {{Einstein Telescope}}},
  author = {Zhao, W. and Broeck, C. Van Den and Baskaran, D. and Li, T. G. F.},
  year = 2011,
  month = jan,
  journal = {Physical Review D},
  volume = {83},
  number = {2},
  eprint = {1009.0206},
  primaryclass = {astro-ph},
  pages = {023005},
  issn = {1550-7998, 1550-2368},
  doi = {10.1103/PhysRevD.83.023005},
  urldate = {2026-01-29},
  archiveprefix = {arXiv}
}

@misc{zhaoLocalizationAccuracyCompact2018,
  title = {Localization Accuracy of Compact Binary Coalescences Detected by the Third-Generation Gravitational-Wave Detectors and Implication for Cosmology},
  author = {Zhao, Wen and Wen, Linqing},
  year = 2018,
  month = mar,
  eprint = {1710.05325},
  primaryclass = {astro-ph},
  doi = {10.1103/PhysRevD.97.064031},
  urldate = {2025-01-01},
  archiveprefix = {arXiv},
  langid = {english}
}

\end{document}